\documentclass[conference]{IEEEtran}
\IEEEoverridecommandlockouts
\usepackage{cite}
\usepackage{amsmath,amssymb,amsfonts}
\usepackage{algorithmic}
\usepackage{graphicx}
\usepackage{textcomp}
\usepackage{xcolor}
\def\BibTeX{{\rm B\kern-.05em{\sc i\kern-.025em b}\kern-.08em
    T\kern-.1667em\lower.7ex\hbox{E}\kern-.125emX}}

\usepackage{ascmac}
\usepackage[final]{listings}
\usepackage{tikz}\usetikzlibrary{cd}
\usepackage{url}
\usepackage{hyperref}
\usepackage{subfig}
\usepackage{footmisc}
\usepackage{multirow}
\usepackage{moresize}
\usepackage[normalem]{ulem}

\newcommand{\delOp}{\mathsf{delete}}
\newcommand{\insOp}{\mathsf{insert}}
\newcommand{\relOp}{\mathsf{relabel}}
\newcommand{\movOp}{\mathsf{move}}

\begin{document}

\title{On the Correctness of Software Merge
\thanks{\hspace*{-2.25\parindent}This is the final accepted manuscript of the following published paper:\\
A. Mori and M. Hashimoto, ``On the Correctness of Software Merge,'' in Proc. \emph{40th ACM/IEEE Int. Conf. on Autom. Softw. Eng. (ASE)}, 2025, pp. 2338--2349, doi: 10.1109/ASE63991.2025.00193.\\
\copyright~2025 IEEE.  Personal use of this material is permitted.  Permission from IEEE must be obtained for all other uses, in any current or future media, including reprinting/republishing this material for advertising or promotional purposes, creating new collective works, for resale or redistribution to servers or lists, or reuse of any copyrighted component of this work in other works.
}
}

\author{
\IEEEauthorblockN{Akira Mori}
\IEEEauthorblockA{
\textit{National Institute of}\\
\textit{Advanced Industrial Science and Technology}\\
Ikeda, Osaka, JAPAN \\
ORCID: 0000-0002-3122-3931
}\\
\and
\IEEEauthorblockN{Masatomo Hashimoto}
\IEEEauthorblockA{
\textit{Chiba Institute of Technology}\\
Narashino, Chiba, JAPAN \\
ORCID: 0000-0003-2317-3812
}
}

\maketitle

\begin{abstract}
Three-way merge tools play crucial roles in modern software development, where a developer forks a branch to make local modifications and requests it to be merged into the main branch via a "pull request." 
Despite its importance, the task has traditionally been defined in an intuitive manner, and the results of merge tools are often accepted without scrutiny. 
In this paper, we present a new structural merge tool in comparison with existing tools based on the syntactic criteria we propose for evaluating the merge results.
We require the merge result to be both parsable and universal.
Being parsable means that the result is syntactically valid according to the grammar of the programming language.
Being universal means that the result incorporates all and only the edit operations occurring in each branch while ensuring that edits common to both branches are applied only once.
This requirement can be precisely defined using the notion of pushouts in category theory.
In a large-scale experiment involving 43,774 file merge scenarios from 76 open-source Java projects, we found a number of incorrect results reported by existing tools such as the Git companion merge tool, whereas our tool reports none.
We further compared d3j’s results with 2,582 developer-resolved merges and with 2,459 merge scenarios involving 21 refactoring types.
These experiments revealed both the strengths and current limitations of structural merge, and underscore the importance of clear correctness criteria.
We expect that the proposed criterion will provide a foundation for developing more reliable and principled merge tools.
\end{abstract}

\begin{IEEEkeywords}
  three-way merge, abstract syntax tree (AST), AST comparison, correctness criterion, pushouts, category theory, universal property, partial inclusion map
\end{IEEEkeywords}

\section{Introduction}
\label{sec:intro}
\begin{figure}[b]
  \centering
  \includegraphics[width=.8\linewidth]{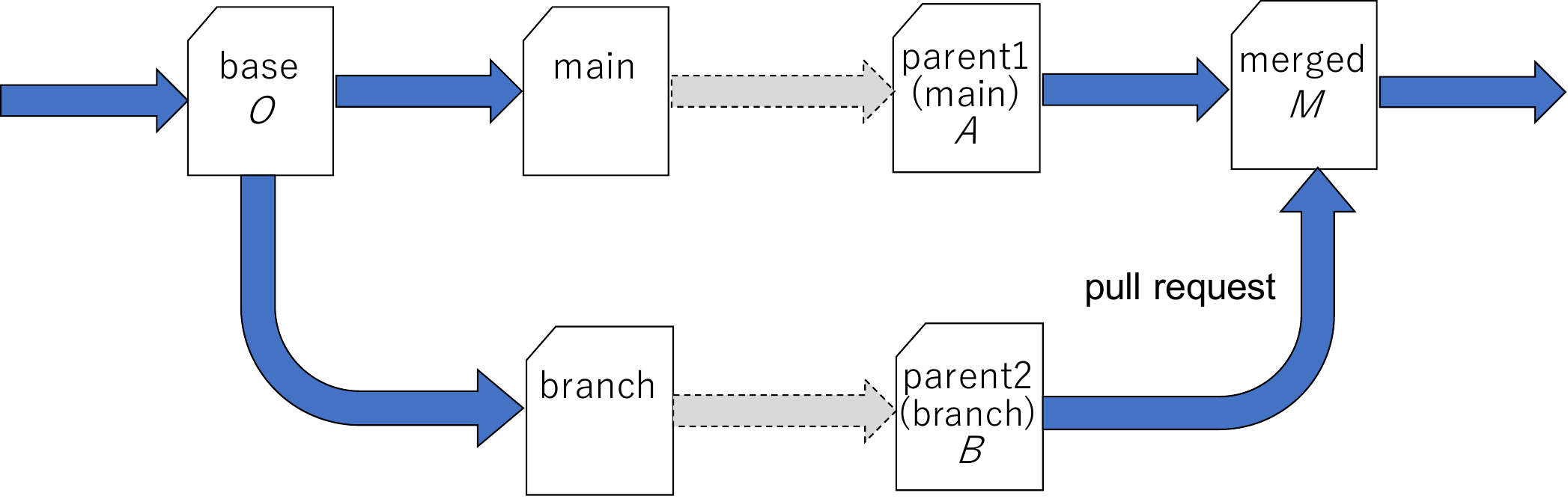}
  \caption{Pull-based software development.}
  \label{fig:merge}
\end{figure}

Three-way program merge is a task to integrate two program versions derived from the same ancestry and has become important due to the widespread use of distributed version control systems (DVCSs).
A DVCS facilitates a distributed workflow in which developers fork branches to work locally and contribute changes by submitting pull requests to project maintainers.
The maintainer processes the ``pull request (PR)'' and incorporates changes into the main branch after evaluating the validity of the contribution.
This so-called pull-based software development model has become a major force since it allows developers to work on new code changes in a more efficient and timely manner~\cite{GousiosEtAl2014}.
Figure~\ref{fig:merge} illustrates this concept.

The task can be complex if the same part of the code is modified in different ways, which may lead to conflicts when divergent changes cannot be resolved.
Various tools have been developed that alert users to conflicts and merge the code when it is safe to do so.
The textual line-oriented tools have been widely used since the diff3 command incorporated three-way merging capability in 1988.
The Git system includes an automated merge tool, which we refer to as git-merge, to streamline tasks within repositories.
However, line-oriented tools like this are limited in their ability to distinguish cosmetic changes, such as extra line breaks and whitespace, and cannot accurately handle fine-grained modifications to specific syntactic elements like variables and statements. 
Consequently, these tools tend to be overly conservative, often flagging conflicts even in straightforward cases.
To address the issue, structural merge tools have been developed, such as JDime~\cite{LessenichEtAl2015}, Spork~\cite{LarsenEtAl2022}, IntelliMerge~\cite{ShenEtAl2019}, 3DM~\cite{Lindholm2004}, and Mastery~\cite{ZhuEtAl2023}.
These tools incorporate fine-grained changes at the level of abstract syntax trees (ASTs) to process version pairs that cannot be merged by the line-oriented tools.

The new development has highlighted the importance of comparing the capabilities of the tools.
Traditionally, tools are evaluated based on the number and size of conflict-free merges they report, but this approach has a significant drawback: it does not account for trivial merges that simply overwrite the entire content, for instance.
Any arbitrary program text could technically be considered a “merge” result, though it would be meaningless in terms of source code integrity.
How, then, can we determine whether one merge result is better than another or even optimal for a given pair of versions?
To answer this, we need robust correctness criteria for evaluating merge results.

This paper proposes to employ the syntactic validity and the universality of entity mappings for evaluating the correctness of merge results.
The former is the grammatical correctness.
In fact, most tools we compare report a number of merge results that are syntactically incorrect.
The latter is a mathematical property of the partial inclusion maps on AST nodes induced by the edit sequences computed between ASTs for the given version pair. 
The property is derived from the concept of pushouts in category theory, for which we examine the commutativity and the optimality for the merge result.
The commutativity requires the two merging routes to represent the same partial map, and the universality requires the result to be the most universal in the sense that any other merge can be obtained by modifying it in a unique way up to isomorphisms.
To our surprise, even git-merge reports a number of merge results that are not universal.

We present our three-way merge tool for Java programs, called d3j, and demonstrate its capabilities by way of a large-scale comparative study involving 7 existing tools over 43,774 per file merge scenarios from 76 open source Java projects.
The analysis shows that the tools in comparison report a number of incorrect merge results due to inappropriate algorithm design\slash implementation or insufficient conflict identification, whereas our tool reports none of them thanks to its ability to mix and reorder edit operations on ASTs and its detailed built-in conflict rules that keep merge results syntactically consistent.
In addition to the large-scale study, we compared d3j against developer-resolved merges, which reveals cases where human choices diverge from formal correctness.
We also conducted a refactoring-centered experiment, showing how different tools handle complex structural edits and why correctness criteria matter in such scenarios.
Considering the importance of the merging task in modern software development, we believe that the study offers useful insights into the development of advanced tools in the future.

The main contributions of the paper are as follows:
\begin{itemize}
\item We propose syntactic correctness criteria for evaluating merge results.
  The criteria are based on the concepts from category theory and can be used to spot incorrect merge results reported by other tools including git-merge.
\item We present our own merge tool called d3j that can serve as a reference implementation.
  For the large datasets used in a comparative study, d3j reports no incorrect merge results.
\item We conduct an extensive comparison with existing open source and research tools.
  The result helps to understand the difference in design and implementation of the algorithms.
\end{itemize}

The rest of the paper is organized as follows.
Section~\ref{sec:backgrounds} motivates our approach, and Section~\ref{sec:overview} reviews methods for AST differencing and merging with explanations of the rules for identifying conflicts.
Section~\ref{sec:criteria} explains the evaluation method using partial inclusion maps of AST nodes.
Section~\ref{sec:experi} details the experiments conducted to show the effectiveness of the tool and the evaluation method in comparison with other tools.
After describing related work in Section~\ref{sec:related}, we conclude the paper in Section~\ref{sec:concl} with explanations of future work.

\section{Backgrounds}
\label{sec:backgrounds}
In this section, we clarify the research context for the paper by explaining the limitations of the textual line-oriented merge tools and the emerging issues with new structural tools.

First, let us define the symbols for describing the merge situations throughout the paper.
Consider a file merge scenario where an original file $O$ is copied into the developer's branch and modified to a file $B$, while in the main repository, $O$ is modified to a file $A$, as shown in Figure~\ref{fig:merge}.
A merge tool outputs a merged file $M$ that incorporates changes from $O$ to $A$ and from $O$ to $B$.
When it cannot find a consistent merge, it reports conflicts.

\begin{figure}[tb]
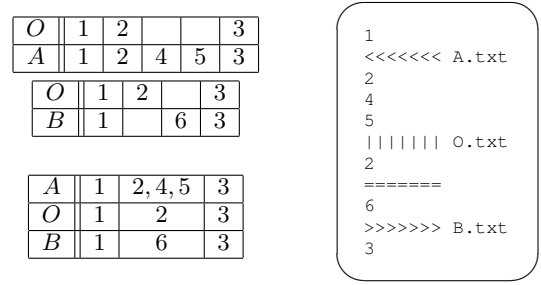

  \small
  \centering
  \begin{minipage}{.4\linewidth}
    \centering
    $\begin{array}{|c||c|c|c|c|c|} \hline
       O & 1 & 2 &   &   & 3  \\ \hline
       A & 1 & 2 & 4 & 5 & 3  \\ \hline
     \end{array}$

     \vspace{1mm}

     $\begin{array}{|c||c|c|c|c|} \hline
        O & 1 & 2 &   & 3 \\ \hline
        B & 1 &   & 6 & 3 \\ \hline
      \end{array}$

      \vspace{5mm}

      $\begin{array}{|c||c|c|c|} \hline
         A & 1 & 2,4,5 & 3 \\ \hline
         O & 1 &   2   & 3 \\ \hline
         B & 1 &   6   & 3 \\ \hline
       \end{array}$
     \end{minipage}\hfil
     \begin{minipage}{.3\linewidth}
       \begin{screen}
         \scriptsize
\begin{verbatim}
1
<<<<<<< A.txt
2
4
5
||||||| O.txt
2
=======
6
>>>>>>> B.txt
3
\end{verbatim}
       \end{screen}
     \end{minipage}
     \caption{The Diff3 algorithm and its output.}
     \label{fig:diff3}
\end{figure}

It may appear that line-oriented tools such as diff3 and git-merge are sufficient for daily software development tasks.
These tools treat files as sequences of symbols, assuming that there is an injective mapping from a set of lines to a set of symbols and employ the longest common subsequence (LCS) algorithm or the minimum edit distance (MED) algorithm to find the optimal matchings between $O$ and $A$ and between $O$ and $B$ as shown in the upper left of Figure~\ref{fig:diff3}, where $O=[1,2,3]$, $A=[1,2,4,5,3]$ and $B=[1,6,3]$.
For simplicity, we use digits for symbols.

According to the analysis by Khanna and others~\cite{Diff3-2007}, the tool overlays the two matchings along the sequence in $O$ to obtain a stack of aligned symbol sequences from files $O$, $A$, and $B$, as shown in the lower left of Figure~\ref{fig:diff3}.
This leads to an output shown in the right side of Figure~\ref{fig:diff3} as a result of the command ``\texttt{diff3 -mA A.txt O.txt B.txt}'', which means that there is a conflict and the user may choose $O$, $A$ or $B$ in order to resolve the conflict.
The git merge produces the same result for this case.

This behavior seems too conservative since $A$ and $B$ can be merged easily.
Let us assume $1$ corresponds to a text "Heat up the oil in the pan.", $2$ to "Stir fry chicken.", $3$ to "Add vegetables.", $4$ to "Add soy sauce.", $5$ to "Add seasoning." and $6$ to "Stir fry pork." in imaginary cooking recipes.
One would naturally expect a recipe "Heat up the oil in the pan. Stir fry pork. Add soy sauce. Add seasoning. Add vegetables.", a sequence $[1,6,4,5,3]$, as the result of the merge, just by changing chicken to pork and also by adding soy sauce and seasoning.
In fact, we see many cases like this when merging pull requests using line-oriented tools.
It would be so much better if these were merged automatically rather than being resolved manually.

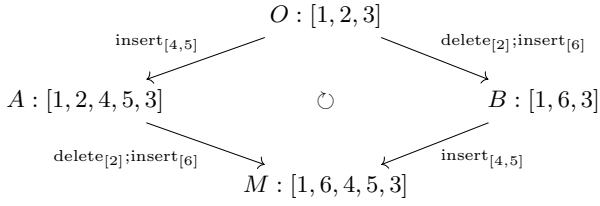
\begin{figure}[tb]
  \centering
  \small
  \begin{tikzcd}
    & O:[1,2,3] \arrow[rd, "\mathrm{delete}_{[2]};\mathrm{insert}_{[6]}"] \arrow[ld, "\mathrm{insert}_{[4,5]}"']& \\
    A:[1,2,4,5,3]  \arrow[rd, "\mathrm{delete}_{[2]};\mathrm{insert}_{[6]}"'] & \circlearrowright & B:[1,6,3] \arrow[ld, "\mathrm{insert}_{[4,5]}"]\\
    & M:[1,6,4,5,3] &
  \end{tikzcd}
  \caption{Merge as commutative diagram.}
  \label{fig:comm}
\end{figure}

The situation becomes clear if we view three-way merging as forming commutative diagrams, as shown in Figure~\ref{fig:comm}, by regarding arrows as edit sequences between program texts.
For instance, the arrow between $O$ and $A$ and between $O$ and $B$ in Figure~\ref{fig:diff3} can be viewed as an edit that inserts $4$ and $5$ in between $2$ and $3$ and an edit that replaces $2$ with $6$ (in other words, deletes $2$ and then inserts $6$), respectively.
Computing a merge $M$ looks straightforward for this case.

Besides being too conservative, the line-oriented tools strongly depend on the ways the program texts are split into lines.
For instance, if the entire code is formatted as a single line, a conflict will arise unless the two versions are identical.
These tools work well only when the texts are formatted in a human-friendly manner, which poses consistency issues when processing machine-generated code.

These concerns motivated the development of many structural merge tools, which brings us back to the issue of the trivial merge when we compare the capabilities of the tools.
For instance, the tool may report an arbitrary content such as $[7,8,9]$ as a merge result for the problem shown
in Figure~\ref{fig:comm} by removing all lines in $A$ and $B$ and inserting arbitrary lines $7$, $8$, and $9$.
On what grounds can we say such a result is useless compared to $[1,6,4,5,3]$, which requires non-trivial computation?
There have been theoretical suggestions that software merge can be regarded as a pushout in category theory~\cite{Goguen1995,MimramEtAl2013,Diaconescu2021}.
Pushouts model optimal amalgamation processes in an abstract setting and fit well with the view of commutative diagrams shown in Figure~\ref{fig:comm}.
We propose correctness criteria based on the pushout construction in the category of sets and partial inclusion maps.
A key observation is that a partial inclusion map $f:A\rightarrow B$ corresponds to an edit from $A$ to $B$ where an element in $A$ that does not have an image in $B$ corresponds to a deleted node and the element in $B$ that does not belong to the image of $f$ to an inserted node.
For example, a partial inclusion map from $O$ to $B$ depicted in Figure~\ref{fig:diff3} represents an edit that deletes 2 in $O$ and inserts 6 in $B$.
A full theory of tree merging further requires a partial order with tree conditions (as proposed by Moerdijk and Oosten~\cite{MoerdijkOosten2018}) and order preservation of maps.
Although we do not delve into the details of theoretical developments in this paper, the correctness criteria we propose provide practical methods for assessing merge tools by identifying redundant or missing elements in their merge results.

\section{Technical Framework}
\label{sec:overview}
\begin{figure}[tb]
  \centering
  \includegraphics[width=\columnwidth,clip]{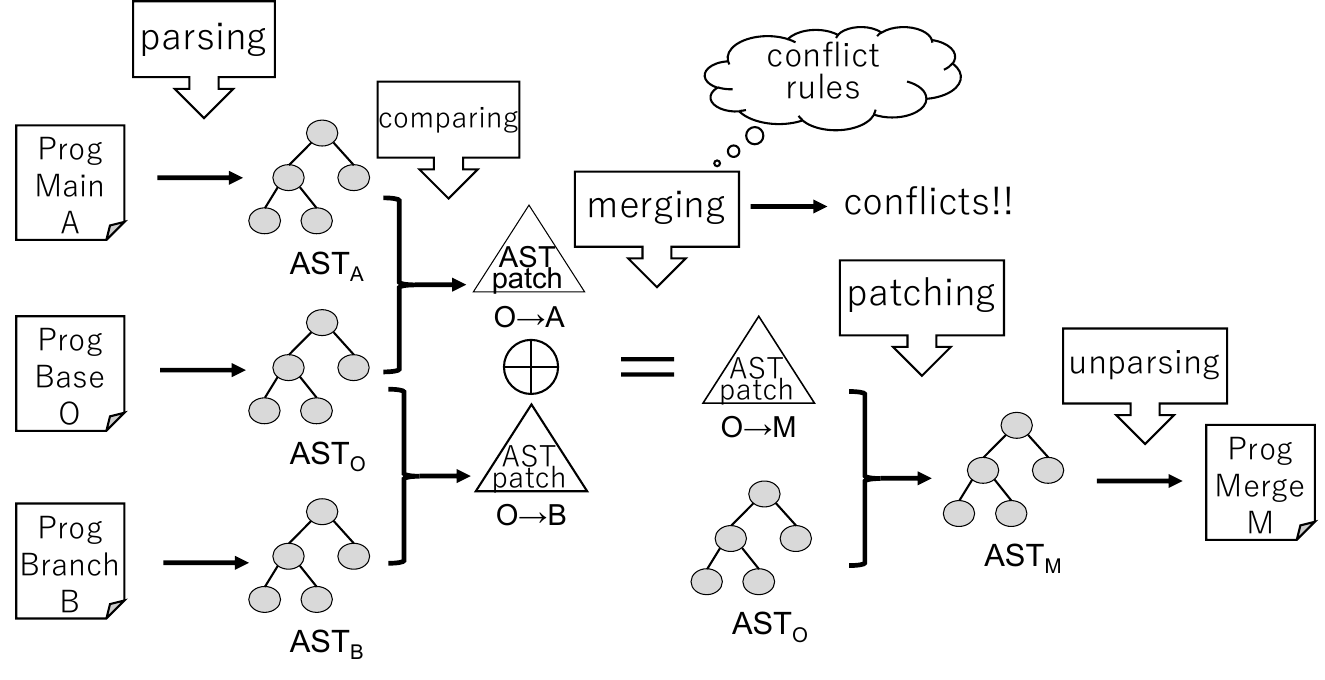}
  \caption{Merge system overview.}
  \label{fig:system}
\end{figure}

We will explain the technical framework of our merge system.
Figure~\ref{fig:system} shows the entire process.
In the following, we explain each component.

\subsection{Parsing}

We developed our own Java parser specifically for AST comparison.
It is implemented in OCaml using an LR(1) parser generator called Menhir~\cite{Menhir}.
It started with ``The Java Language Specification, Third Edition''~\cite{jls3} and now can process Java 11 programs.
It emits an AST that faithfully represents the parsing process.
An output AST has named nodes exclusively for named entities such as variables, methods, and classes to restrain the application of the relabel operation.

\subsection{AST Comparison}
\label{sec:astdiff}

At the core of every merge tool is a comparison algorithm that calculates the matching of entities between given pairs of program texts, and the merging is performed by integrating matching along $O\rightarrow A$ and $O\rightarrow B$.
For a line-oriented merge tool, a string comparison algorithm is used to calculate line-by-line matching~\cite{Gusfield1997}, and for a structured merge tool, a tree comparison algorithm is used to calculate node-by-node matching~\cite{Bille2005}.
As we stated earlier, these matchings can be viewed as partial inclusion maps between sets of entities.
For program text comparison, we require inclusion maps to be order preserving on the grounds that insertion and deletion should not alter the order of the program.
For the line-oriented edit model, the linear order is assumed and for the AST based edit model, the tree order, which is the order of the ancestry and sibling relations, is assumed.

\begin{figure}[tb]
  \centering
  \includegraphics[width=.8\columnwidth, clip]{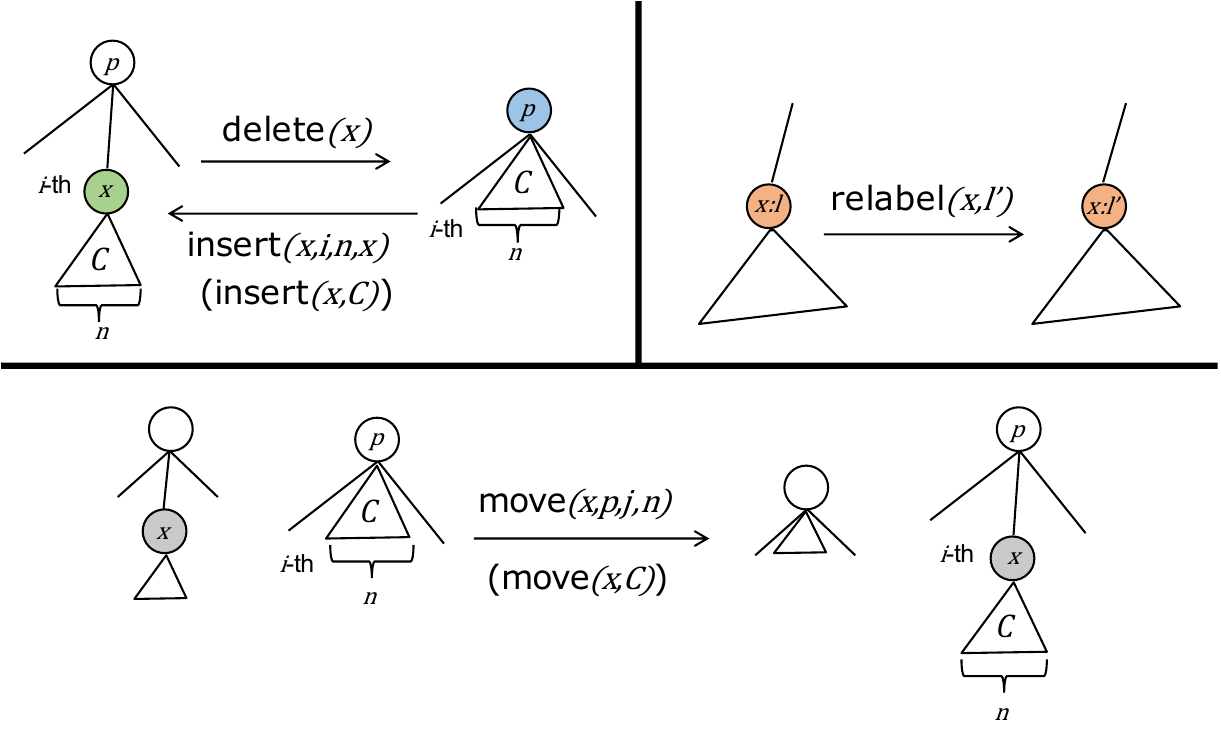}
  \caption{Tree edit operations.}
  \label{fig:editops}
\end{figure}

We briefly explain the tree comparison method underlying our merge tool.
The AST comparison algorithm computes for a given pair of ASTs $T_1$ and $T_2$ a sequence of edit operations that transforms $T_1$ into $T_2$.
We consider four edit operations: $\delOp$, $\insOp$, $\relOp$, and $\movOp$,  each operating on an AST node.
Figure~\ref{fig:editops} illustrates these edit operations.
Note that we use colors consistently throughout the paper, that is, green for nodes to be deleted, blue for nodes to be inserted, orange for nodes to be relabeled, and gray for nodes to be moved.

We use an approximated version of the optimal algorithm proposed by Zhang and Shasha~\cite{ZhangShasha1989}, incorporating language-specific features and post-processing for move detection and relabel reorganization.
The algorithm keeps track of partial inclusion maps between $T_1$ nodes and $T_2$ nodes by regarding a move\slash relabel operation as a combination of delete operations and insert operations.

\subsection{AST Merging}
\label{sec:patch}

\begin{figure}[tb]
  \centering
  \includegraphics[width=.6\linewidth]{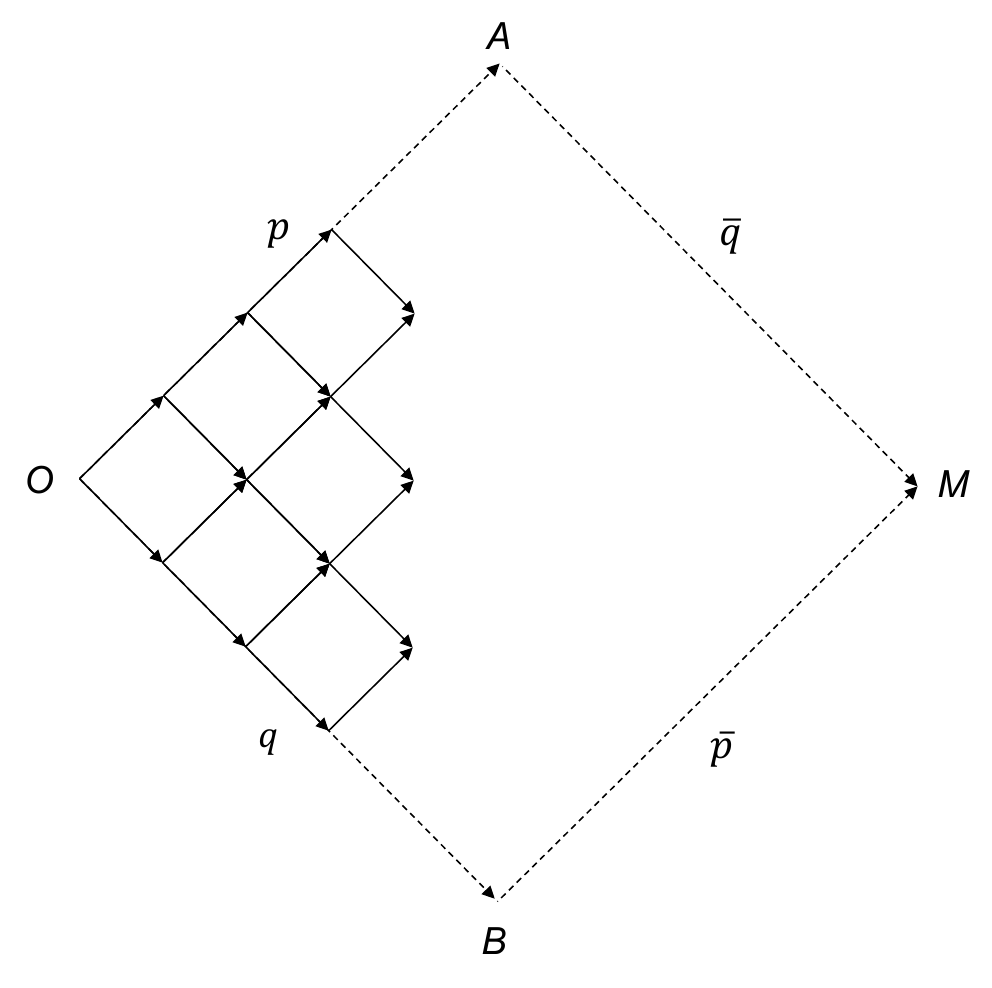}
  \caption{Completing commutative diagram of edits.}
  \label{fig:completion}
\end{figure}

A merge task can be reformulated as calculating a pair of AST matchings $\overline{q}:A\rightarrow M$ and $\overline{p}:B\rightarrow M$ that form a commutative diagram for a given merge scenario $p:O\rightarrow A$ and $q:O\rightarrow B$.
This is achieved by successively constructing a commutative diagram made of a single edit operation until the common target $M$ is obtained as shown in Figure~\ref{fig:completion}.

The objective is to compute an optimal commutative diagram in which the merge result $M$ incorporates all and only edit operations from $p$ and $q$, ensuring that any edits common to $p$ and $q$ are applied only once.
In category-theoretic terms, this is equivalent to computing a pushout for $p$ and $q$.
However, software merging involves more than simply constructing a commutative diagram, as the merge result must also be syntactically valid.

In order to maintain syntactic consistency, a merge tool enforces its own conflict rules that prohibit certain pairs of edit Operations from being merged.

\subsection{Rules of Conflicts}
\label{sec:conflicts}

The conflict rules depend on the grammar of the target programming language, making them challenging to generalize theoretically.
By carefully examining the grammar of the Java language, we established 32 conflict rules, implemented in approximately 170 Python functions and totaling over 6,000 lines of code.
Typically, a conflict rule is defined for a pair of edit operations, allowing the check to be performed during the diagram completion step, as illustrated in Figure~\ref{fig:completion}.

\begin{figure}[tb]
  \centering
  \includegraphics[width=.995\linewidth]{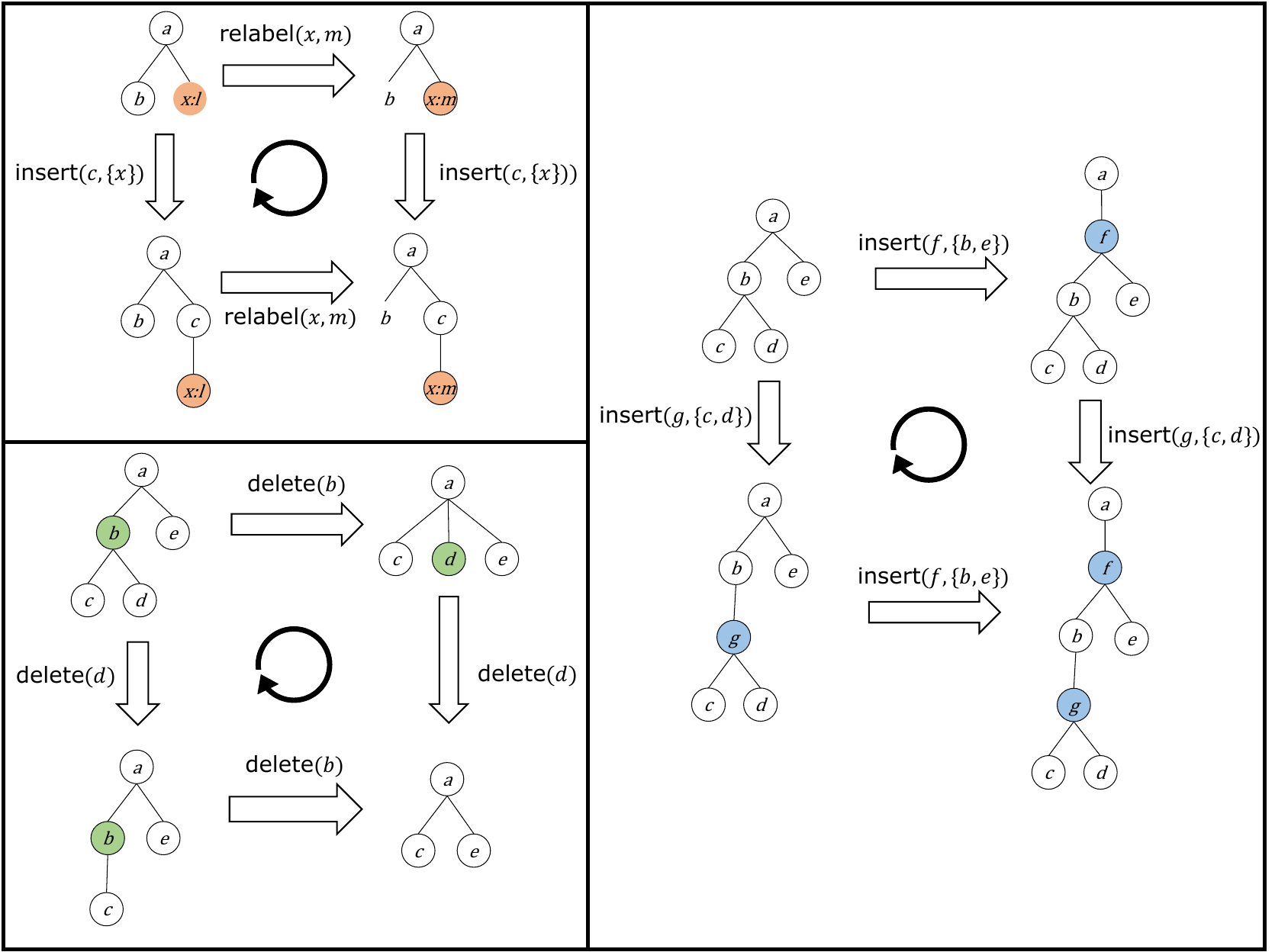}\\
  \includegraphics[width=.99\linewidth]{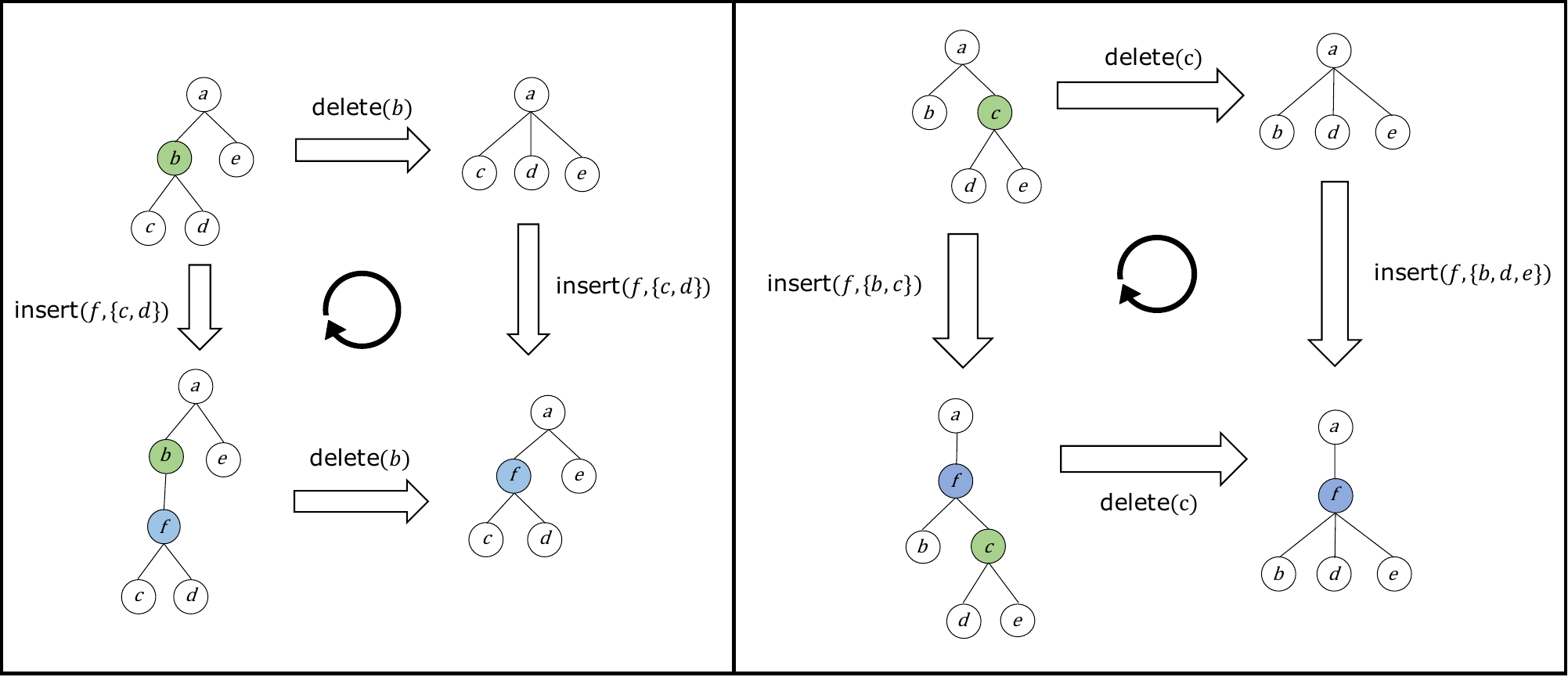}
  \caption{Commutativity between tree edit operations.}
  \label{fig:commutativity}
\end{figure}

We explain some important rules below.

\begin{list}{$\bullet$}{}

\item $\relOp$-$\relOp$ (name clash):
A conflict is reported when a named entity is relabeled in two different ways such as a variable $a$ relabeled to $b$ in $A$ and to $c$ in $B$.

\item $\delOp$-$\delOp$ (split deletion): A conflict is reported when $A$ and $B$ delete connected groups of nodes that do not coincide but overlap.
If we ignore programming contexts, we could merge $A$ and $B$ by deleting all nodes that are going to be deleted in $A$ and $B$.
However, as a program, the parts not in the common area of deletion are likely to cause syntactic inconsistencies.
Suppose $A$ changes an expression ``\texttt{f(c)}'' to ``\texttt{x=c}'' by first deleting ``\texttt{f(\_)}'' and then inserting ``\texttt{x=\_}'' and $B$ deletes the entire expression ``\texttt{f(c)}''.
If we merge them as trees, we get a partial expression ``\texttt{x=}'' that does not parse.
We would delete overlapped segments only when they coincide as in the case where both $A$ and $B$ delete an entire expression ``\texttt{f(c)}''.

\item $\insOp$-$\delOp$ (broken dependency):
A conflict is reported if a parent node $p$ of an inserted node in $A$ gets deleted in $B$ when no stable ancestors of $p$ belong to the same syntactic category as $p$ does.
The inserted node is likely to get lost if we delete the parent node in $M$.
For instance, we should not mix in $M$ the deletion of a method declaration (in $B$) and the insertion of a statement in its body (in $A$). But on the other hand, we can mix in $M$ the deletion of an inner class $c$ (in $B$) and the insertion of a method in its body (in $A$) as long as the outer class of $c$ is not deleted.

\item $\insOp$-$\insOp$ (cardinality mismatch):
An insert operation in $A$ conflicts with another in $B$ if both get under the same parent at the same sibling slot but introduce different numbers of child nodes beneath them.
There is a high possibility of syntactic inconsistencies.
Suppose a ``\texttt{switch}''-statement is inserted in $A$ and an ``\texttt{if}''-statement is inserted in $B$ at the same spot.
These two insertion operations may not commute because a ``\texttt{switch}''-statement tends to have more child nodes than an ``\texttt{if}''-statement.
It is therefore safer to flag a conflict than to commit the merge.

\item $\relOp$-$\delOp$ (name inconsistency):
A conflict is reported when a named entity such as a method declaration is relabeled in $A$ but is deleted in $B$.
The named entity may appear in many other places such as method invocations hence there is a high chance that relabeling in $A$ and deletion in $B$ do not match in the def-use chain of the entity.
Named entities such as annotations do not impose a def-use chain and hence may cause unexpected side effects.
The risk is too high to commit the merge overall.

\item Syntactic consistencies:
Our tool takes advantage of its syntax awareness to maintain the validity of the merge results by checking the syntactic categories of the nodes upon insertion and deletion.
\begin{itemize}
  \item When the parent node admits a variable number of child nodes, a new child node can be inserted if its syntactic category is consistent with that of the parent node.  Any child node can be deleted.
  \item When the parent node admits a fixed number of child nodes, a child node can be inserted\slash deleted if the syntactic categories of all child nodes become compatible after insertion\slash deletion, respectively.
\end{itemize}

\end{list}

\subsection{AST patching}

The result of AST comparison is stored in an XML file as a patch for updating the AST $O$.
Our patch files contain relative path information for each edit operation, they can be applied to ASTs other than the ones used for computing it.

\subsection{Unparsing}

When the merge is complete, the tool presents the result in a textual format by traversing the AST in depth-first order.

This process of reconstructing source code text from an AST is commonly called \emph{unparsing}. 
In some literature it is also referred to as \emph{pretty printing}, although strictly speaking, pretty printing usually emphasizes formatting and layout, while unparsing refers to systematically generating code text from the AST.

For the purposes of this paper, information about textual layouts and comments in $O$, $A$, and $B$ are not recorded in the corresponding ASTs, since they are not relevant for evaluating and comparing tools.
Note that it is not difficult to record the positions of comments and restore them after the merge.
Comments can also be merged if necessary.

\section{Correctness Criteria for Merge Results}
\label{sec:criteria}
We now explain the correctness criteria we propose for evaluating merge results.

In category theory, the pushout of arrows $f:O\rightarrow A$ and $g:O\rightarrow B$ consists of an object $M$ and two arrows $i_1:A\rightarrow M$ and $i_2:B\rightarrow M$ such that:
\begin{itemize}
  \item  $i_1\circ f = i_2\circ g$ (commutativity),
  \item for any other such tuple ($N$, $j_1$, $j_2$) satisfying the above commutativity, there exists a unique arrow $u:M\rightarrow N$ such that $j_1=u\circ i_1$ and $j_2=u\circ i_2$ (universality).
\end{itemize}
The configuration is illustrated in Figure~\ref{fig:pushout}.
A pushout, when it exists, is unique up to isomorphism.
A reader unfamiliar with category theory may view an arrow as a function and an object as a set.
In the category of sets and functions, a pushout is formed by taking the disjoint union \( M \) of \( A \) and \( B \) and identifying \( f(a) \) with \( g(a) \) in \( M \) for all \( a \in O \).

\begin{figure}[tb]
  \centering
  \small
  \begin{tikzcd}
    & O \ar[rd, "g"] \ar[ld, "f"']& \\
    A \ar[rd, "i_1"'] \ar[dddr, "j_1"', bend right=20] \ar[dddr, phantom, "\circlearrowright"] & \circlearrowright &
    B \ar[ld, "i_2"] \ar[dddl, "j_2", bend left=20] \ar[dddl, phantom, "\circlearrowright"] \\
    & M \ar[dd, "u"] & \\
    & &  \\
    & N &
  \end{tikzcd}
  \caption{Pushout.}
  \label{fig:pushout}
\end{figure}

In general, a pushout represents the most general way of combining structures and has been proposed as a theoretical framework for studying software merging~\cite{Goguen1995,MimramEtAl2013}.
Category theory provides an abstract and unified framework for understanding mathematical structures and relationships.
In our case, pushouts in the category of ordered sets and partial inclusion maps help us analyze the problem of software merging.

We view a node matching computed by the tree comparison algorithm as a partial inclusion map between node sets and define the correctness criteria by the concept of pushouts in the category of partial inclusion maps.
Mathematically, a partial inclusion map $f:A\rightarrow B$ is a partial function such that $f(a)=a$ whenever $a\in A$ and has an image in $B$.
Since the composition of partial inclusions is also a partial inclusion, they form a category.

\begin{figure}[tb]
  \centering
  \includegraphics[width=\columnwidth,clip]{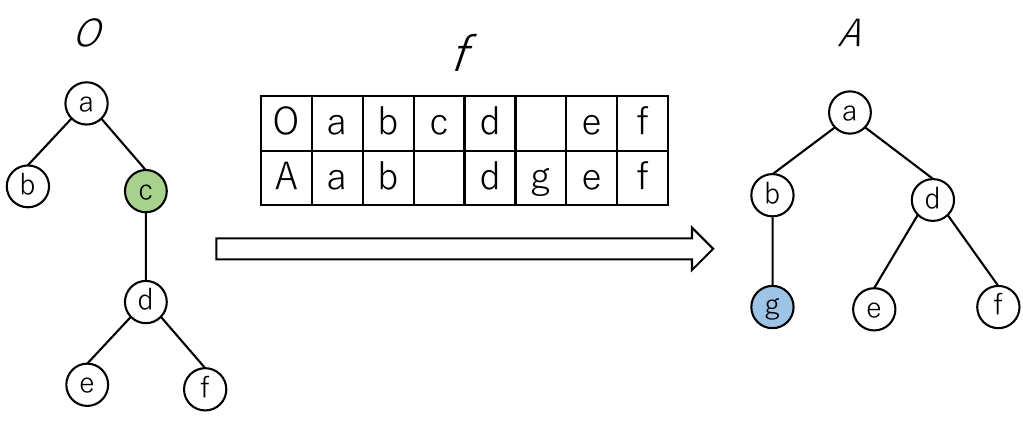}
  \caption{Edit sequence as order-preserving partial inclusion map.}
  \label{fig:partial}
\end{figure}

Consider a tree edit shown in Figure~\ref{fig:partial} for example.
The edit consists of insertion of node $g$ and deletion of node $c$.
One can see that the partial inclusion map $f$ associated with the edit preserves the tree order; i.e., $x\leq y$ in $O$ implies $f(y)\leq g(y)$ in $A$ for $a,b,d,e,f$, and that the element of $O$ that does not have an image in $A$ corresponds to the deleted node and the element in $A$ that does not belong to the image of $f$ to the inserted node.

In software merge scenarios, we naturally expect that the underlying partial inclusion maps form a pushout diagram, as anticipated in Figure~\ref{fig:comm}.
Formally, for a software merge $f:O\rightarrow A$, $g:O\rightarrow B$, $i_1:A\rightarrow M$ and $i_2:B\rightarrow M$, we require the following conditions\footnote{The simplification assumes that node identities are globally unique, which can be achieved via $\alpha$-renaming when modeling edits as partial inclusions.}.
\begin{enumerate}
  \item\label{cond:junk} For any node $x\in M$, either $x\in A$ or $x\in B$. In other words, no redundant node should be added.
  We call this the {\bf no extra insertion} condition.
  \item\label{cond:insdel} Any node inserted in either $A$ or $B$ must have an image of $i_1$ or $i_2$ in $M$, respectively.
  In other words, no inserted node should be omitted from the result.
  We call this the {\bf no missed insertion} condition.
  \item\label{cond:commondel} Any node that has images in both $A$ and $B$ also has an image in $M$.
  In other words, a node that is not deleted should remain. We call this the {\bf no extra deletion} condition.
  \item\label{cond:nmdel} Any node that does \emph{not} have an image in either $A$ or $B$   also does \emph{not} have an image in $M$. 
  In other words, a node deleted in either $A$ or $B$ must not reappear in $M$. 
  We call this the {\bf no missed deletion} condition.
\end{enumerate}

As a reminder, we explain the concepts of equality and composition for partial functions.
The equality of partial functions differs from that of total functions; two partial functions are equal if and only if they are defined for the same set of elements and have the same values wherever they are defined.
The composition of two partial functions \( f: A \rightarrow B \) and \( g: B \rightarrow C \), denoted \( g \circ f \), is defined as a new partial function from \( A \) to \( C \) such that 
$(g \circ f)(a) = g(f(a))$ if both $f(a)$ and $g(f(a))$ are defined and undefined otherwise.
 Note that for \( g \circ f \) to be defined at a point \( a \in A \), both \( f(a) \) and \( g(f(a)) \) must be defined.

\begin{figure}[tb]
  \centering
  \includegraphics[width=.8\columnwidth,clip]{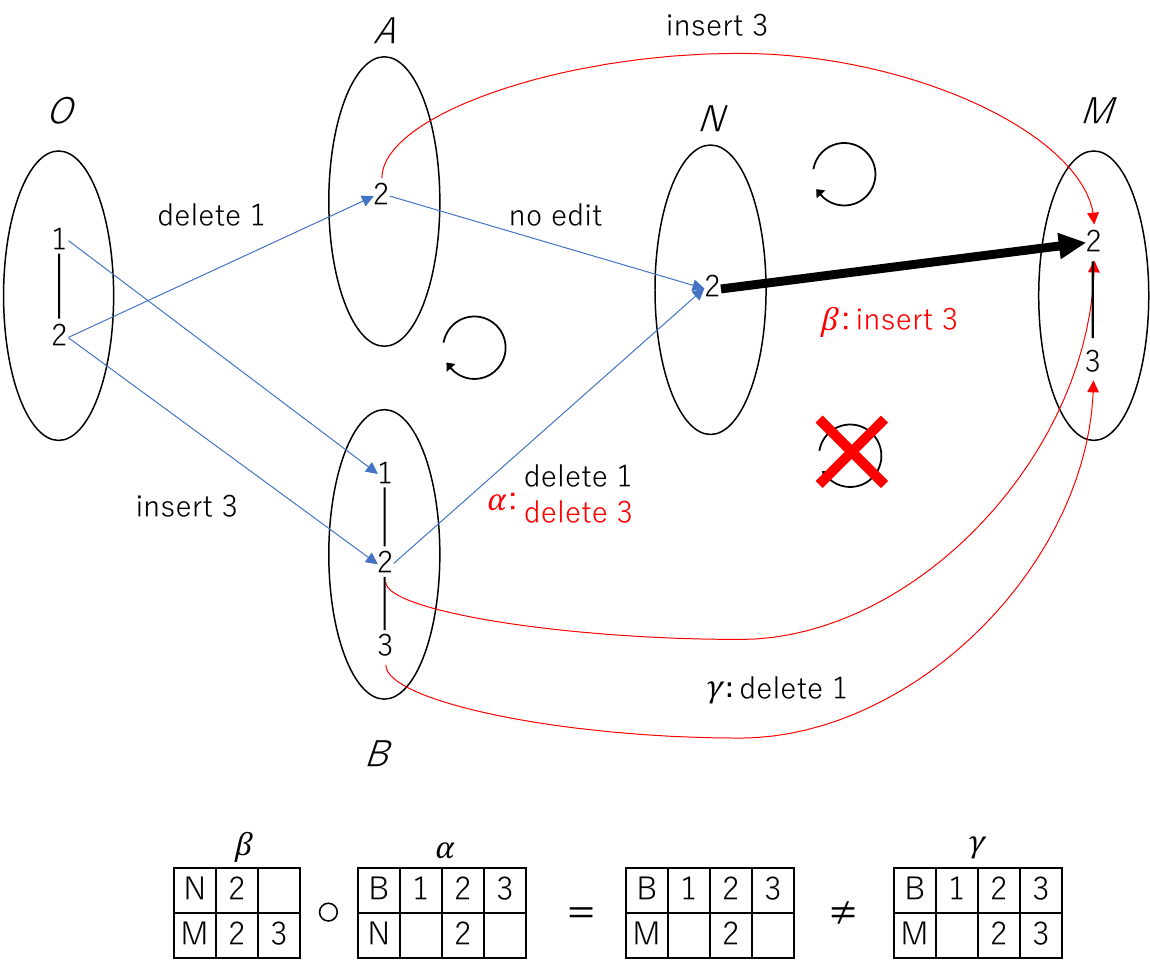}
  \caption{
  An example where deleting inserted nodes breaks universality.
  $M$ represents a correct merge and $N$ represents an incorrect merge that deletes the inserted node $3$.
  $M$ is correct since it satisfies all conditions.
  $N$ satisfies the commutativity condition, however, it does not satisfy the {\bf no missed insertion} condition
  and there exists no arrow from $N$ to $M$ satisfying the commutativity for universality unless $N$ retains node $3$.
  Clearly, $M$ is a better merge result than $N$.
  }
  \label{fig:insdel}
\end{figure}

These explain why a trivial merge is not considered correct.
If an element is deleted in the original code, an optimal merge retaining that element will never be reached.
A concrete example illustrating this is shown in Figure~\ref{fig:insdel}.
In fact, the {\bf no missed insertion} condition and the {\bf no extra deletion} condition are intended to reject trivial and less general merge results.

One may argue that these conditions apply to sets of nodes but do not obviously extend to trees.
We can define a category $\mathcal{T}$ of tree edits by viewing a tree as a partially ordered set that has a least element and is such that ancestors of an element are well-ordered (i.e., linearly ordered in finite cases) as proposed by Moerdijk and Oosten~\cite{MoerdijkOosten2018} and by defining a morphism between trees as a partial inclusion map.
It is straightforward to prove that if $\mathcal{T}$ has a pushout, then it remains a pushout in the category of sets with partial inclusion maps when the ordering is disregarded.
We developed our own merge tool d3j together with a tool to check the conditions~\ref{cond:junk}-\ref{cond:nmdel} using our own AST comparison tool.
In fact, the merging algorithm implemented in d3j is illustrated in Figures 6 and 7 and is designed to calculate pushouts while checking for conflicts.

Since the commutativity violations are very rare and often covered by the other conditions, we focus on the universality conditions \ref{cond:junk}-\ref{cond:commondel} for the rest of the paper.
In addition to the universality, we require the merge result to be syntactically consistent; i.e., parsable.

It is important to emphasize that in our study, correctness is defined in a formal sense. 
A conflict-free merge is considered correct if and only if it is both parsable and universal. 
This definition is entirely mathematical and does not rely on external annotations or human judgment. 
In other words, instead of comparing against a manually constructed ground truth, we check whether the merge result satisfies these two well-defined properties.

\section{Experiments}
\label{sec:experi}
We design three experiments targeting complementary aspects of merge evaluation: (i) scalability and correctness on a large dataset, (ii) comparison against human developer choices, and (iii) stress testing under refactoring-heavy scenarios.

\subsection{Tools in Comparison}
\label{sec:tools}

We chose git-merge, imediff~\cite{Imediff2003}, jFSTMerge~\cite{CavalcantiEtAl2017}, JDime~\cite{LessenichEtAl2015}, IntelliMerge~\cite{ShenEtAl2019}, AutoMerge-PTM~\cite{ZhuEtAl2019}, Mastery~\cite{ZhuEtAl2023} and  Spork~\cite{LarsenEtAl2022} for evaluation and comparison.

\textbf{Git-merge} is a line-oriented merge tool used in the Git system.
Although its implementation details are undocumented, it is known that it started as a drop-in replacement of the RCS Merge tools.
It has been enhanced with various heuristics for handling program texts such as ``indent heuristics.''\footnote{\url{https://lkml.org/lkml/2017/8/4/481}}
We used version:2.25.1 for the experiments.

\textbf{Imediff} is an open-source tool that was originally developed by Jarno Elonen in 2003 in Python\footnote{\url{https://github.com/osamuaoki/imediff}}.
It is a line-oriented tool that attempts to mix edit operations.
It has a unique two-stage strategy, where line-by-line matching is attempted in the first stage and then character-by-character matching is tried with the same algorithm.
We used commit:d5d0b4a for the experiments.

\textbf{jFSTMerge} is a research product developed by Cavalcanti and others~\cite{CavalcantiEtAl2017} in an effort to verify and improve a method for semi-structured merge studied by Apel and others~\cite{ApelEtAl2011}.
The original study aimed at enhancing the capabilities of traditional merge tools such as diff3 by working on tree structures called feature structure trees extracted from various dependencies found in program texts such as module hierarchies.
A simplified tree comparison algorithm is used to match different versions and the so-called tree amalgamation algorithm is used for incorporating changes in two versions.
The amalgamation algorithm extends the longest common subsequence algorithm to tree structures.
jFSTMerge is a new efficient implementation of the original tool FSTMerge with several heuristics to help reduce false positives\slash negatives, concerning mapping between renamed entities, type ambiguity, and initialization blocks.
We used commit:e212b39 for the experiments.

\textbf{JDime} is an AST based structured merge tool developed by Le{\ss}enich and others in an attempt to improve the capabilities of FSTMerge~\cite{LessenichEtAl2015} that they previously developed.
JDime is the first merge tool based on ASTs and employs the same amalgamation algorithm as FSTMerge.
JDime uses a meta-compilation system called JastAddJ\footnote{\url{http://jastadd.org/web/}} for parsing the Java programs.
We used commit:63ffc34 for the experiments.

\textbf{IntelliMerge} is a graph-based merge tool developed by Shen and others in an attempt to overcome the difficulties in merging the code involving refactoring~\cite{ShenEtAl2019}.
It converts program versions into graphs and performs merging by aligning the nodes considering the potential refactoring involving them.
We used v1.0.9 for the experiments.

\textbf{AutoMerge-PTM} is an AST based structural merge tool developed by Zhu and others in an attempt to enhance the capability of AutoMerge with precise tree matching~\cite{ZhuEtAl2019}.
AutoMerge is an interactive merge tool developed by the same group on top of JDime by formulating an idea of exchanging modifications made in two versions as a formal system called a version space algebra~\cite{ZhuEtAl2018}.
We used commit:e73038b for the experiments.

\textbf{Spork} is an AST based structured merge tool developed by Larsen and others.
It builds on an AST comparison tool called GumTree~\cite{FalleriEtAl2014} and incorporates a three-way merge tool for XML called 3DM~\cite{Lindholm2004}.
GumTree is based on the edit model proposed by Chawathe and others~\cite{ChawatheEtAl97} with focus on detection of move operations.
Spork relies on a software analysis library called Spoon~\cite{Spoon} for parsing Java programs.
We used commit:22ddb0f for the experiments.

\textbf{Mastery} is an AST based structural merge tool developed by the same group as AutoMerge-PTM~\cite{ZhuEtAl2023}.
Mastery improves the accuracy in amalgamating the ASTs of $A$ and $B$ by spotting the so-called shifted code among them.
We used commit:defa08c for the experiments.

\subsection{Large-Scale Merge Experiment}
\label{sec:large-exp}

This experiment evaluates correctness and scalability of d3j on a wide dataset of open-source Java projects.
For the experiments in this paper, we used a server with a dual 128-core x86\_64 processor, a 2.7GHz base frequency, 1536GB of memory, and running Ubuntu Linux 24.04. We ran the tool on the dataset with 240 processes.

The results are shown in Table~\ref{tab:result}.
Counts of conflict free merges (CFMs), counts of CFMs that match the merged code in the repositories (called the perfect CFMs), counts of CFMs that are not syntactically valid (well parsed), counts of CFMs that violate the universality condition,
the cumulative processing time (in minutes),
and finally, counts of correct CFMs, our key performance indicator,
are shown for each tool.

\begin{table}
  \centering
  \resizebox{\columnwidth}{!}{%
    \begin{tabular}{crrrrrr}
    \hline
    Tool & CFM\# & perfect-CFM\# & non-WP\# & non-Univ\# & Time (min.) & CCFM\# \\
    \hline
    git-merge     & 31,533 & 31,099 &     0 &    40  &   2.1   & 31,493 \\
    imediff       & 36,752 & 34,096 &   196 &   130  &  51.5   & 36,426 \\
    jfstmerge     & 31,680 & 29,233 &    76 &   171  & 2137.0  & 31,433 \\
    jdime         & 36,992 &  7,716 &    89 &   128  & 9495.3  & 36,775 \\
    automergeptm  & 38,586 &  7,564 &   597 & 1,080  & 10534.7 & 36,909 \\
    intellimerge  & 37,541 & 20,293 & 19,565 & 5,265 &  599.6  & 12,711 \\
    spork         & 39,361 &  8,012 &   213 & 1,573  & 4447.7  & 37,575 \\
    mastery       & 39,098 & 33,766 &     9 &   270  & 1314.2  & 38,819 \\
    d3j           & 39,541 & 36,015 &     0 &     0  &  863.1  & 39,541 \\
    \hline
    \end{tabular}
  }
  \caption{Large-scale merge experiment results.}
  \label{tab:result}
\end{table}

Note that JDime and Spork modify the input source file without undoing it in the output.
Typically, JDime transforms an \texttt{if}-statement ``\texttt{if (...) \{...\} else if \{...\}}'' to ``\texttt{if (...) \{...\} else \{if \{...\}\}}'' and Spork transforms an array instance ``\texttt{\{0,1,2,3\}}'' to ``\texttt{new int[]\{0,1,2,3\}}''.
We identified about 20 such patterns and implemented the undoing functions in our Java parser.
Without this, the numbers of perfect CFMs would have been much lower for these tools.

The breakdown of universality violations is shown in Table~\ref{tab:univ}.
The rightmost column of the table contains the numbers of cases where d3j reports conflicts, for which we do not know yet if there exist correct merges.
Note that the classification is not exclusive.
There are overlaps between types of universality violations.

\begin{table}
  \centering
  \resizebox{\columnwidth}{!}{%
  \begin{tabular}{ccccc|c}
\hline
Tool &
Extra insert &
Missed insert &
Extra delete &
Missed delete &
d3j conflicts \\
\hline
\multirow{2}{*}{git-merge} 
 & 92.50\% & 5.00\% & 0.00\% & 2.50\% & 22.50\% \\
 & (37/40) & (2/40) & (0/40) & (1/40) & (9/40) \\
\hline
\multirow{2}{*}{imediff} 
 & 78.46\% & 15.38\% & 0.00\% & 11.54\% & 29.23\% \\
 & (102/130) & (20/130) & (0/130) & (15/130) & (38/130) \\
\hline
\multirow{2}{*}{jfstmerge} 
 & 53.22\% & 46.78\% & 1.17\% & 2.92\% & 5.85\% \\
 & (91/171) & (80/171) & (2/171) & (5/171) & (10/171) \\
\hline
\multirow{2}{*}{jdime} 
 & 7.81\% & 83.59\% & 32.03\% & 7.81\% & 14.84\% \\
 & (10/128) & (107/128) & (41/128) & (10/128) & (19/128) \\
\hline
\multirow{2}{*}{automergeptm} 
 & 13.52\% & 92.69\% & 3.80\% & 1.39\% & 20.93\% \\
 & (146/1080) & (1001/1080) & (41/1080) & (15/1080) & (226/1080) \\
\hline
\multirow{2}{*}{intellimerge} 
 & 18.88\% & 4.52\% & 0.04\% & 84.01\% & 5.11\% \\
 & (994/5265) & (238/5265) & (2/5265) & (4423/5265) & (269/5265) \\
\hline
\multirow{2}{*}{spork} 
 & 46.79\% & 53.40\% & 10.30\% & 2.23\% & 17.48\% \\
 & (736/1573) & (840/1573) & (162/1573) & (35/1573) & (275/1573) \\
\hline
\multirow{2}{*}{mastery} 
 & 19.26\% & 75.19\% & 1.48\% & 4.81\% & 27.78\% \\
 & (52/270) & (203/270) & (4/270) & (13/270) & (75/270) \\
\hline
\end{tabular}}
  \caption{Types of universality violations across tools.}
  \label{tab:univ}
\end{table}

\begin{figure}[tb]
  \centering
  \includegraphics[width=.9\linewidth]{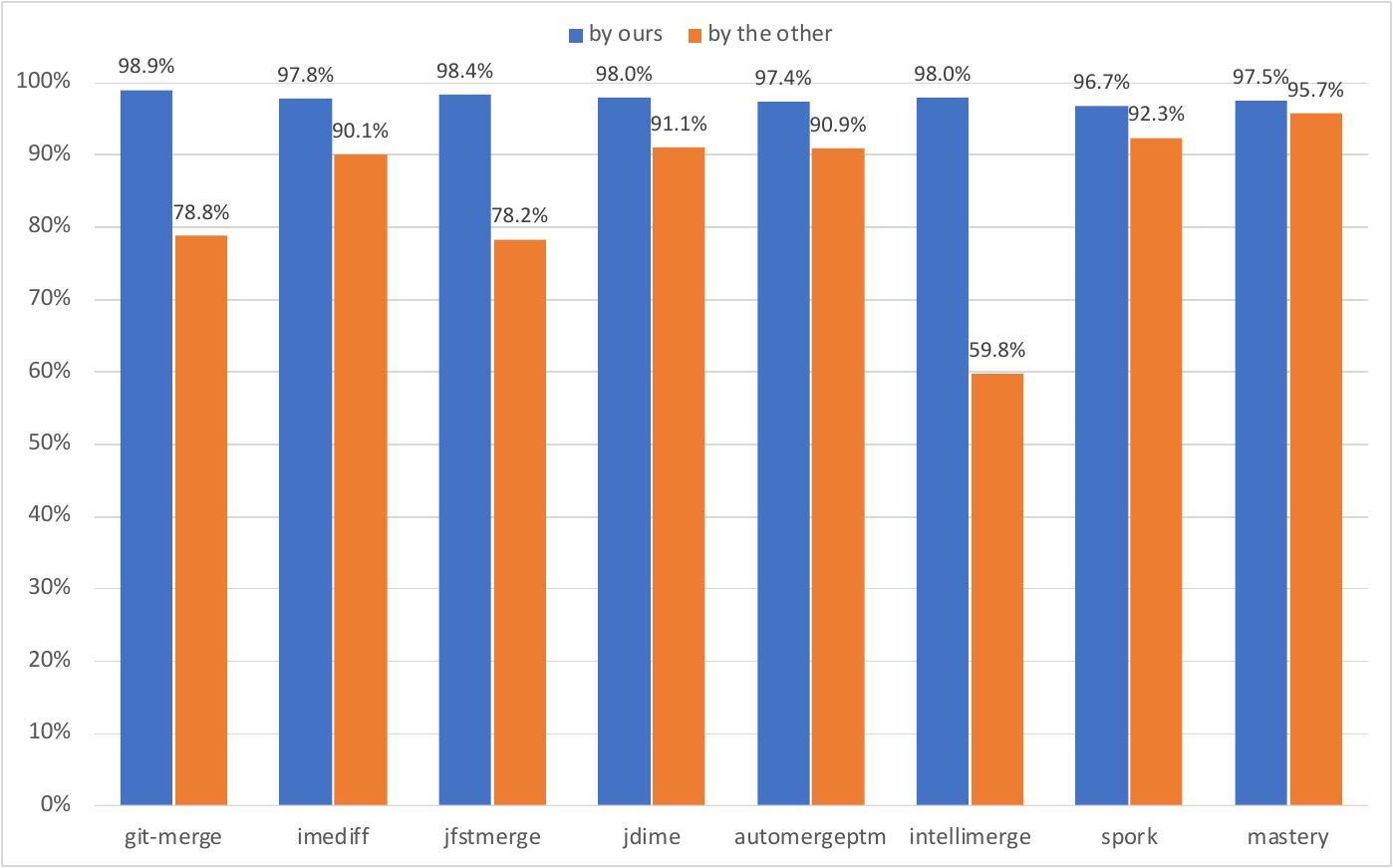}
  \caption{Mutual coverage of correct conflict-free merges.}
  \label{fig:coverage}
\end{figure}

For pairwise comparison of the tools, the mutual coverage of correct merge scenarios is shown for d3j.
Let $\mathrm{CCFM}(T)$ denote a set of merge scenarios that are correctly merged by tool $T$.
Let $W$ denote our tool d3j.
For each tool $A$ other than $W$, a pair of bars representing ratios $|\mathrm{CCFM}(A)\cap\mathrm{CCFM}(W)|/|\mathrm{CCFM}(A)|$ (mutual coverage by d3j) and $|\mathrm{CCFM}(A)\cap\mathrm{CCFM}(W)|/|\mathrm{CCFM}(W)|$ (mutual coverage by other tool $A$), where $|S|$ denotes the size of a set $S$, are shown in Figure~\ref{fig:coverage}.

\begin{figure*}[tb]
  \centering
  \includegraphics[width=.9\linewidth]{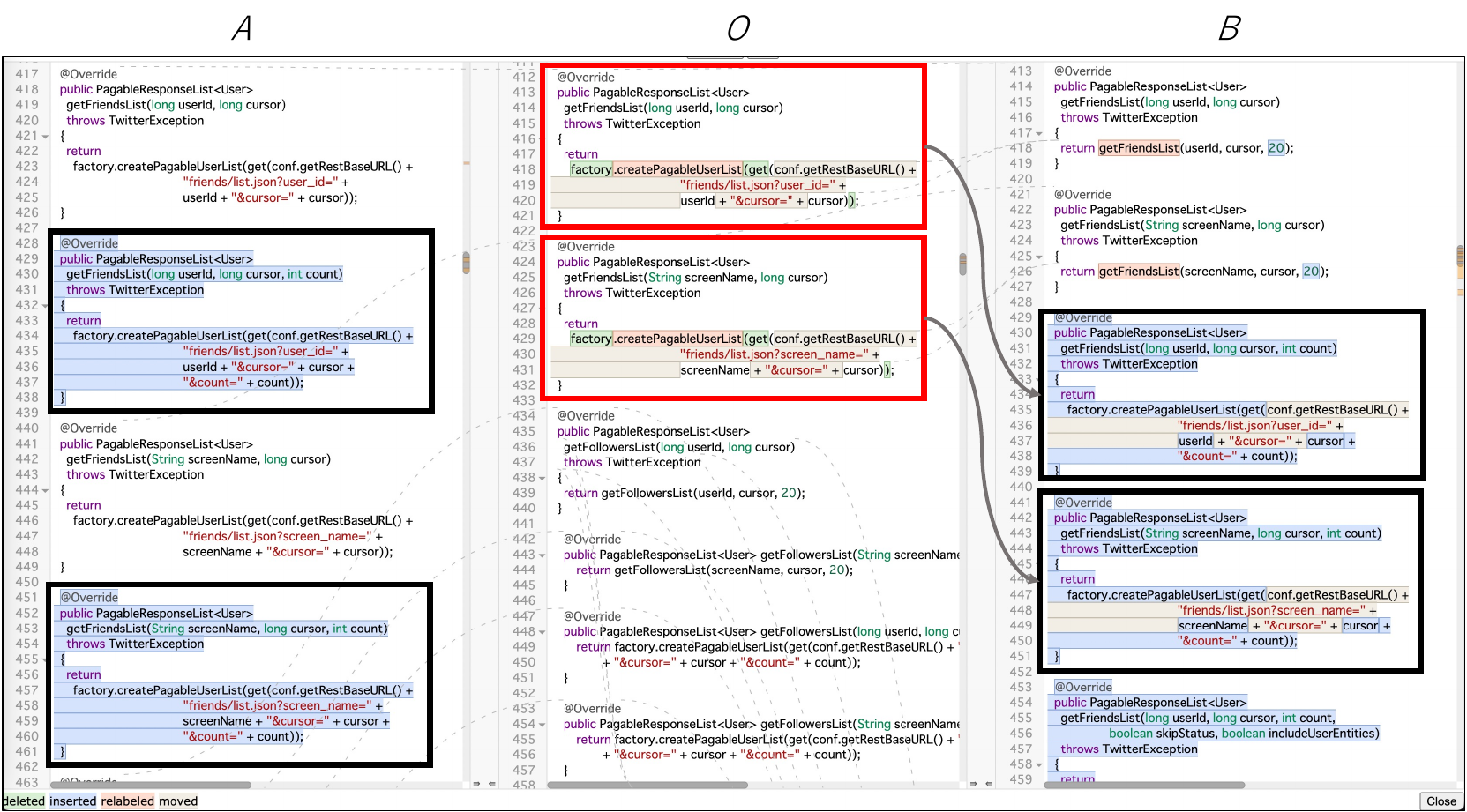}
  \caption{A merge scenario that caused an incorrect CFM reported by git-merge. The code segments in the red boxes in $O$ are left unchanged in $A$ but modified in $B$ to coincidentally match the newly inserted code segments in $A$ (in the black boxes). An insert-insert conflict should have been reported since the insert in $A$ and the insert in $B$ collides.  However, git-merge reports a non-universal CFM that has duplicated copies of the code in the black boxes.}
  \label{fig:incorrect1}
\end{figure*}

We summarize the analysis of errors for each tool.
\begin{list}{$\bullet$}{}
\item Git-merge reported no syntactic errors but had a few universality violations.
  It suffers in rare cases when the same syntactic entities appear simultaneously in $A$ and $B$.
  Figure~\ref{fig:incorrect1} shows such an example.

\item Imdediff is fast and successfully covers many cases.
  However, it emits syntactically wrong results when it fails in the character-by-character matching phase, for which 
  there is much room for improvement.

\item JDime, jFSTMerge, and AutoMerge-PTM share the same tree amalgamation algorithm, which fundamentally relies on accurate tree matching between $A$ and $B$. Failures in identifying correct node matchings likely resulted in the high counts of violations of both the no extra insertion and no missed insertion conditions.
The statistics show many such failures, which lead to merge results M containing redundant elements or missing inserted nodes.
The low number of d3j conflicts in these cases suggests that these are actual errors, since d3j returns correct results where the other tools fail.

\item IntelliMerge reports a high number of syntactically incorrect merge results due to duplicated entities.
We are afraid that the tool is in an early development phase and may require further work for improvements.

\item Mastery shows the signs of improvements in its AST merger. However, a relatively high number of universality violations suggests that it still has a trait of the amalgamation algorithm of JDime and others.

\item Spork shows a relatively high number of universality violations related to the {\bf no missed insertion} condition.
This may be caused by the merge engine called 3DM that Spork employs.
It is documented that 3DM cannot detect conflicts where a delete operation in one version overlaps with an edit operation, particularly an insertion, in another, resulting in a typical case as shown in Figure~\ref{fig:insdel} of Section~\ref{sec:criteria}.

\item d3j demonstrates comparable capability in computing conflict-free merge results against Spork and Mastery.
Among these, d3j is the only tool without any syntactic or universality violations.
One of its few drawbacks is processing speed; among AST-based tools, d3j is outperformed by IntelliMerge in this regard.

\end{list}

A merge scenario that caused an incorrect CFM reported by git-merge is shown in Figure~\ref{fig:incorrect1}.
Other tools including Spork and Mastery report numerous incorrect CFMs that drop an inserted code block, violating the universality condition as illustrated by an example shown in Figure~\ref{fig:insdel}.

Our guarantees rely on accurate AST differencing. Misidentifications (e.g., renames vs. insertions) can cause incorrect or overly conservative merges. Existing tools like GumTree often produce redundant edits, while compiler parsers abort on incomplete code, limiting their suitability for merge automation.

\subsection{Comparison with Human Resolutions}
\label{sec:human-exp}

This experiment evaluates how d3j compares against actual developer choices in repositories.
We used the dataset by Miraldo~\cite{Miraldo2020}, which contains 2,582 Java merge scenarios where git-merge reported conflicts and the developers manually committed resolutions. By analyzing these developer-chosen results alongside d3j’s outputs, we examine how often human resolutions satisfy our correctness criteria, and whether d3j can resolve conflicts differently.

Of the 2,582 scenarios, d3j reported 1,652 correct conflict-free merges, whereas 711 merge results recorded in the repositories were judged incorrect according to our correctness criteria.
Fifteen were syntactically invalid, such as cases where conflict markers were left unerased.
Of the 696 merge scenarios in which the human-resolved results violated the universal property, 328 were successfully merged by d3j without conflicts.
Note that only syntactically valid results $M$ can be checked for universality, since it requires computing edit sequences from $A$ and $B$ to $M$.

\begin{figure*}[tb]
  \centering
  \includegraphics[width=\linewidth]{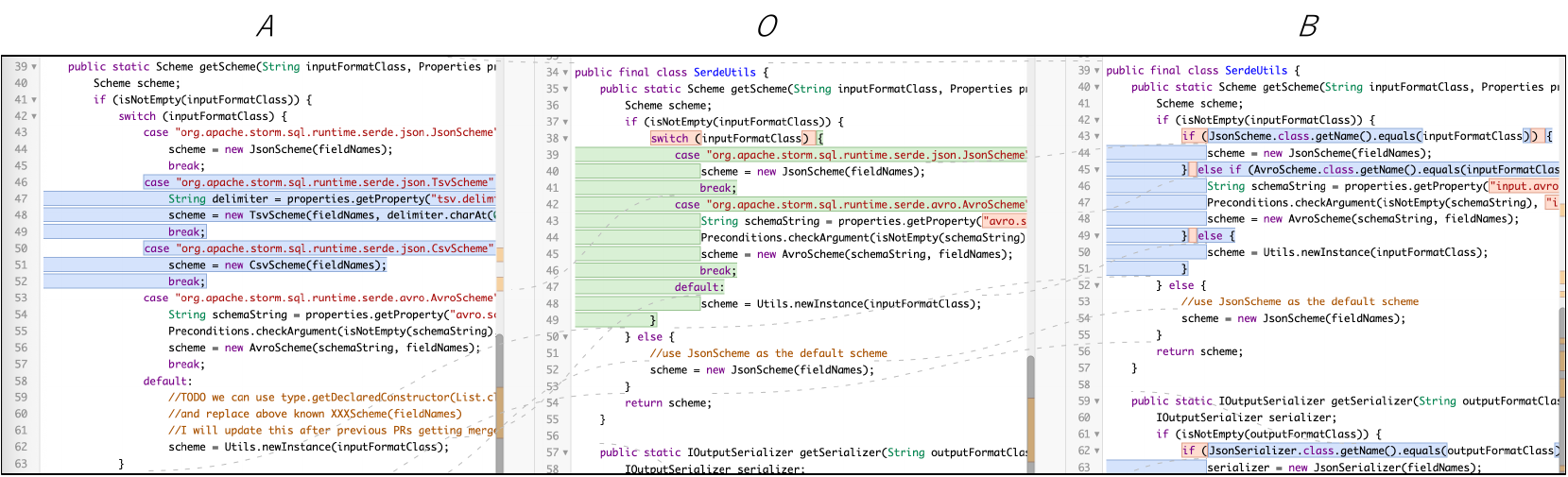}
  \caption{Example of semantic merge resolved by a human (switch $\rightarrow$ if refactoring).
}
  \label{fig:semantic-merge}
\end{figure*}

We randomly selected 50 scenarios from each group and manually inspected how humans resolved the conflicts.
All d3j conflict-free but repository-incorrect cases were confirmed as instances where developers added new code after computing correct merges.
Among d3j conflict and repository-incorrect cases, all but one were situations where the developers picked the changes made in either $A$ or $B$, exclusively or simultaneously.
Only one semantic merge was found: a case where new \texttt{case} clauses were added to a {\texttt switch} statement in $A$ while in $B$ the \texttt{switch} statement was converted to a series of \texttt{if} statements.
Figure~\ref{fig:semantic-merge} shows $O$, $A$, and $B$ for this example.
The merge result $M$ should be clear from the context.

By manually inspecting 100 randomly selected cases, we confirmed that our merge result checker worked as expected according to our correctness criteria.
The result is consistent with previous studies~\cite{GhiottoEtAl2018,DeepMerge2023}, which claimed that the majority of human conflict resolutions are performed by choosing existing code from $O$, $A$, or $B$.
However, to our surprise, there were very few cases where humans performed a semantic resolution beyond the capabilities of automated tools.

Considering that humans "correctly" merged at least 1,871 scenarios (2,582 minus 711), while d3j managed 1,652, humans are more capable of resolving git-conflicted merge scenarios.
The gap arises from the difficulty of defining a fixed set of conflict rules that ensure the syntactic validity of results.
Most of our rules are context dependent, meaning that a rule may be ignored in certain cases.
Some conflict rules are deliberately conservative; while this increases reported conflicts, it prevents silent incorrect merges and reflects a trade-off between coverage and correctness.
Continuous fine-tuning of the rules is therefore necessary for further improvements.

Finally, we give some thoughts on why humans often choose suboptimal merges.
The most common cases are when developers are forced to take either $A$ or $B$ and drop the changes in the other.
This violates universality, but it is almost automatic and, as earlier studies showed, very common in practice.

Another frequent pattern is when developers look as if they are aiming for the correct merge, but in the process they replicate or extend changes beyond what merging requires.
They may notice a useful change in $A$ or $B$ and decide to spread it, even though it was not part of the original edits.
These additions count as extra insertions and lead to universality violations.

Such extra edits may be acceptable when humans resolve conflicts, but they are not permissible for automated tools.
In the end, we found very few cases of humans creating genuinely new semantic integrations.
Most of the “suboptimal” merges are either forced choices or opportunistic edits.

\subsection{Refactoring-Centered Experiment}
\label{sec:refactoring-exp}

This experiment stresses merge tools on structurally challenging refactoring scenarios.
We extracted 2,459 conflict-inducing merges across 20 GitHub projects, identified with the help of the tool developed by Mahmoudi and others~\cite{MahmoudiEtAl2019}, which further utilizes a tool called RefactoringMiner developed by Tsantalis and others~\cite{TsantalisEtAl2022} for automatically identifying refactorings.

The resulting dataset covers over 2,000 refactorings of 21 different types, ranging from frequent cases such as Extract Method and Rename Method to rarer patterns like Move \& Rename Class. This experiment evaluates how tools handle merges where structural edits dominate.

Table~\ref{tab:refactoring} shows the result of the experiment.

The experiment poses a challenge for merge tools, as the rate of conflict-free merges is much lower than in the previous experiments.
The strong performance of Spork and d3j is not surprising, since both tools rely on high-performance AST differencing.
IntelliMerge, on the other hand, did not perform as well as expected.
The main issue seems to be its way of mapping nodes, as pointed out by Ellis and others~\cite{EllisEtAl2023}.
It uses similarity heuristics on names, signatures, and local context, but these often fail to capture the correct correspondences in real merges.
When the mapping goes wrong, the tool ends up either mismatching elements or giving up and reporting conflicts.

The two-stage strategy of Imediff has proven highly effective in this experiment.
In terms of the key performance indicator CCFM (correct conflict-free merges), it matches the results of some of the more advanced structural merge tools.

Finally, we touch on the trade-off between the CFM rates and the number of incorrect merges by taking a closer look at Spork.
As noted earlier, Spork is known to ignore delete-change conflicts.
If a code block is deleted in $A$, then any edits to that block in $B$ are simply dropped, even if they were newly inserted elements.
This behavior produces many universality violations, and we observe a high number of such errors in our experiments.

Still, if you only count the raw number of conflict-free merges, Spork looks strong.
This is because in many refactoring cases, dropping edits from $B$ still leaves compilable code, so the tool seems to succeed more often than it really does.
This shows why counting only conflict-free merges is misleading.
Without correctness criteria, it is easy to be fooled into thinking a tool is performing well when in fact it is discarding valid changes.

\begin{table}
  \centering
    \resizebox{\columnwidth}{!}{%
    \begin{tabular}{crrrrrr}
    \hline
    Tool & CFM\# & perfect-CFM\# & non-WP\# & non-Univ\# & Time (min.) & CCFM\# \\
    \hline
    imediff       &  881 &  462 &   47 &    17 &    3.4  &  817 \\
    jfstmerge     &  266 &  154 &    5 &     4 &   74.6  &  257 \\
    jdime         &  697 &   87 &    2 &    17 &  381.1  &  678 \\
    automergeptm  &  941 &   84 &   77 &   105 &  395.1  &  759 \\
    intellimerge  &  686 &  132 &  430 &   126 &   90.1  &  130 \\
    spork         & 1,309 &   81 &  14 &   232 &  247.4  & 1063 \\
    mastery       &  908 &  465 &    3 &    32 &  111.7  &  873 \\
    d3j           & 1,150 &  615 &   0 &     0 &   74.0  & 1150 \\
    \hline
  \end{tabular}
    }
    \caption{Results of refactoring-centered merge experiment.}
  \label{tab:refactoring}
\end{table}

\subsection{Threats to Validity}
\label{sec:threats}

We manually inspected merge results during development to refine our conflict rules.
We also randomly sampled 100 cases from specific categories of merges to confirm that our checker behaved as expected according to our correctness criteria.
While this sampling gave us confidence, we did not perform a systematic large-scale manual validation, which remains a possible next step for mitigating false positives.

The AST comparison is a complex task, and sometimes even a human cannot decide what happened between the original version and the modified version.
This inevitably indicates that we may have false positives or false negatives, which we did not check due to the volume and complexity of the task.
The purpose of the current paper is to formally present the notion of the correctness of merge and highlight the importance of such criteria.

Although our AST comparison tool employs a generic optimal algorithm developed by Zhang and Shasha~\cite{ZhangShasha1989}, there is a chance that other tools would produce different partial inclusion maps than the ones used for the evaluation.
The fact that each tool uses a different parser suggests that the tools may have worked on ASTs in different shapes.
This adds uncertainty to the identification of the root cause of the incorrect merge.
We know, by manual inspection, that all incorrect results are indeed incorrect from the programming viewpoints.
However, we do not know what caused the incorrect merge.
The merge system is complex and any component illustrated in Figure~\ref{fig:system} can go wrong.
A modular architecture will help since it allows us to have a uniform environment where merge engines are evaluated separately.

The results reported by the merge tool are per file.
There is a chance that some parts of the file are merged without conflicts and other parts encounter conflicts, in which case the tool reports conflicts for the file.
Although we do not believe that the rate of incorrect merge results fluctuates to the granularity of measurement, we may miss instances that provide us with useful insights.
Creating fine-grained datasets helps in this regard but takes much effort.

\section{Related Work}
\label{sec:related}
A survey by Tom Mens contains pointers to early studies in the field~\cite{Mens2002Survey}.
Westfechtel and Buffenbarger studied merge problems in terms of abstract syntactic structures.
Westfechtel proposed a context-sensitive merge in which dataflow augmented syntactic representation is employed~\cite{Westfechtel1991}.
Buffenbarger~\cite{Buffenbarger1995} coined the term of syntactic software merging and proposed a syntax directed merge techniques~\cite{Buffenbarger1995}.
The rise of pull-based software development as described by Gousios and others~\cite{GousiosEtAl2014} stimulated studies on enhanced merge tools.

Our tool and approach differ from the tools listed in Section~\ref{sec:tools} by using AST patch files computed by an AST differencing tool for modifying ASTs.
Employing an edit distance algorithm developed by Zhang and Shasha~\cite{ZhangShasha1989} helped in implementing the ideas of merging changes since renaming is identified as part of the problem and move operations can be reconstructed in post-processing steps.

Mehdi and others proposed a method of evaluating quality of merge by consulting merged file in the repository~\cite{MehdiEtAl2014}.
The idea is similar to the perfect CFMs we measured in Section~\ref{sec:large-exp}.
Zhu and others took the same evaluation method~\cite{ZhuEtAl2023}.
However, such a measure is not appropriate for evaluating the capabilities of the tool since merge commits often contain the changes unrelated to the merging tasks such as bug\slash typo fixes.
We believe that the number of CFMs and the number of incorrect CFMs are good indicators of the tool's capabilities.

Asenov and others considered a way of extending the capabilities of line-oriented merge tools by utilizing unique node IDs provided by tree differencing~\cite{AsenovEtAl2017}.
Although they reported an experimental result using Gumtree~\cite{FalleriEtAl2014}, they have
not released an implementation of the proposed functionality.
Their method may be based on ideas similar to ours.

In this paper, we only considered syntactic conflicts.
Sousa and others defined the notion of conflict-freedom as a semantic requirement for a correct program merge and proposed a method for verifying the property based on lightweight summarization of shared code segments and a unique four-way AST differencing~\cite{SousaEtAl2018}.
Dias and others proposed a method for calculating an impact of change from base
version to versions to be merged by way of change impact analysis.
If the difference of the two impacts, called the delta-impact, is non-empty, then there is a chance of semantic conflicts~\cite{DiasEtAl2015}.
Nguyen and others proposed a method for detecting semantic conflicts by symbolically executing a program that incorporates all changes guarded by ``\texttt{if}''-statements to check an inconsistent program behaviors~\cite{NguyenEtAl2015}.
Since these methods rely on techniques for comparing program structures, applying full-scale AST differencing tools may be useful for studying semantic conflicts.
There are several such tools reported in recent years~\cite{FalleriEtAl2014,HashimotoMori2008,FluriEtAl2007}.

Finally, we mention a category-theoretic study of patches by Mimram and Di Giusto~\cite{MimramEtAl2013}.
They model version control categorically, that is, files as objects, patches as morphisms, and 3-way merge as a pushout.
Since pushouts can fail in the raw setting (e.g., line-based edits), they compute merges in the free finite-colimit completion where pushouts always exist and conflicts correspond to non-existence in the base.
Our approach differs in that we remain within partial inclusion maps on ASTs, avoiding the need for categorical completions.

\section{Conclusion}
\label{sec:concl}
Across large-scale, developer, and refactoring experiments, d3j consistently avoided incorrect merges and highlighted the importance of formal correctness criteria. 

Our findings also suggest why humans often accept suboptimal merges.
In many developer-resolved cases, developers simply took one branch and discarded the other, or opportunistically replicated edits not part of the original changes.
Both patterns violate universality.
While such resolutions may be acceptable in practice, they underline the importance of correctness criteria, since automated tools must not silently discard or duplicate edits. 

By enforcing correctness even when this increases reported conflicts, 
d3j complements human practice and provides a principled baseline for future tool development. 
The proposed correctness criteria are the first of their kind and will support the design of more reliable merge tools.
However, we do not view d3j as a drop-in replacement for git-merge.
It can complement existing tools.

Extending d3j to other languages mainly requires defining compatible parsers and language-specific conflict rules; we have already developed parsers for C/C++, with conflict rules as the next step.
We emphasize that our correctness guarantees concern syntactic and structural validity.
We plan to explore semantic or behavioral correctness in the future.
On a more practical side, we aim to analyze the conflict rules of existing merge tools to improve the capability of d3j and to explore the possibility of tool-independent definitions of conflict rules.

\section{Data Availability}

We have uploaded a replication package at \href{https://doi.org/10.5281/zenodo.13335352}{\nolinkurl{https://doi.org/10.5281/zenodo.13335352}}.
We will include the link to the source code for the entire tool set upon publication of the paper.

\clearpage

\bibliographystyle{IEEEtran}

\begin{thebibliography}{10}
\providecommand{\url}[1]{#1}
\csname url@samestyle\endcsname
\providecommand{\newblock}{\relax}
\providecommand{\bibinfo}[2]{#2}
\providecommand{\BIBentrySTDinterwordspacing}{\spaceskip=0pt\relax}
\providecommand{\BIBentryALTinterwordstretchfactor}{4}
\providecommand{\BIBentryALTinterwordspacing}{\spaceskip=\fontdimen2\font plus
\BIBentryALTinterwordstretchfactor\fontdimen3\font minus
  \fontdimen4\font\relax}
\providecommand{\BIBforeignlanguage}[2]{{%
\expandafter\ifx\csname l@#1\endcsname\relax
\typeout{** WARNING: IEEEtran.bst: No hyphenation pattern has been}%
\typeout{** loaded for the language `#1'. Using the pattern for}%
\typeout{** the default language instead.}%
\else
\language=\csname l@#1\endcsname
\fi
#2}}
\providecommand{\BIBdecl}{\relax}
\BIBdecl

\bibitem{GousiosEtAl2014}
G.~Gousios, M.~Pinzger, and A.~v. Deursen, ``An exploratory study of the
  pull-based software development model,'' in \emph{Proceedings of the 36th
  International Conference on Software Engineering}, ser. ICSE.\hskip 1em plus
  0.5em minus 0.4em\relax New York, NY, USA: ACM, 2014, pp. 345--355.

\bibitem{LessenichEtAl2015}
\BIBentryALTinterwordspacing
O.~Le{\ss}enich, S.~Apel, and C.~Lengauer, ``Balancing precision and
  performance in structured merge,'' \emph{Automated Software Engineering},
  vol.~22, no.~3, pp. 367--397, Sep 2015. [Online]. Available:
  \url{https://doi.org/10.1007/s10515-014-0151-5}
\BIBentrySTDinterwordspacing

\bibitem{LarsenEtAl2022}
S.~Larsen, J.-R. Falleri, B.~Baudry, and M.~Monperrus, ``Spork: Structured
  merge for java with formatting preservation,'' \emph{IEEE Transactions on
  Software Engineering}, pp. 1--1, 2022.

\bibitem{ShenEtAl2019}
\BIBentryALTinterwordspacing
B.~Shen, W.~Zhang, H.~Zhao, G.~Liang, Z.~Jin, and Q.~Wang, ``Intellimerge: A
  refactoring-aware software merging technique,'' \emph{Proc. ACM Program.
  Lang.}, vol.~3, no. OOPSLA, oct 2019. [Online]. Available:
  \url{https://doi.org/10.1145/3360596}
\BIBentrySTDinterwordspacing

\bibitem{Lindholm2004}
\BIBentryALTinterwordspacing
T.~Lindholm, ``A three-way merge for {XML} documents,'' in \emph{Proceedings of
  the 2004 ACM Symposium on Document Engineering}, ser. DocEng '04.\hskip 1em
  plus 0.5em minus 0.4em\relax New York, NY, USA: ACM, 2004, pp. 1--10.
  [Online]. Available: \url{http://doi.acm.org/10.1145/1030397.1030399}
\BIBentrySTDinterwordspacing

\bibitem{ZhuEtAl2023}
\BIBentryALTinterwordspacing
F.~Zhu, X.~Xie, D.~Feng, N.~Meng, and F.~He, ``On the methodology of three-way
  structured merge in version control systems: Top-down, bottom-up, or both,''
  \emph{Journal of Systems Architecture}, vol. 145, p. 103011, 2023. [Online].
  Available:
  \url{https://www.sciencedirect.com/science/article/pii/S138376212300190X}
\BIBentrySTDinterwordspacing

\bibitem{Diff3-2007}
S.~Khanna, K.~Kunal, and B.~C. Pierce, ``A formal investigation of diff3,'' in
  \emph{FSTTCS 2007: Foundations of Software Technology and Theoretical
  Computer Science}, V.~Arvind and S.~Prasad, Eds.\hskip 1em plus 0.5em minus
  0.4em\relax Berlin, Heidelberg: Springer Berlin Heidelberg, 2007, pp.
  485--496.

\bibitem{Goguen1995}
J.~A. Goguen, ``Categorical approaches to merging software changes,'' 1995,
  unpublished draft.

\bibitem{MimramEtAl2013}
\BIBentryALTinterwordspacing
S.~Mimram and C.~Di~Giusto, ``A categorical theory of patches,''
  \emph{Electron. Notes Theor. Comput. Sci.}, vol. 298, pp. 283--307, Nov.
  2013. [Online]. Available:
  \url{http://dx.doi.org/10.1016/j.entcs.2013.09.018}
\BIBentrySTDinterwordspacing

\bibitem{Diaconescu2021}
\BIBentryALTinterwordspacing
R.~Diaconescu, \emph{Implicit Partiality of Signature Morphisms in Institution
  Theory}.\hskip 1em plus 0.5em minus 0.4em\relax Cham: Springer International
  Publishing, 2021, pp. 81--123. [Online]. Available:
  \url{https://doi.org/10.1007/978-3-030-64187-0_4}
\BIBentrySTDinterwordspacing

\bibitem{MoerdijkOosten2018}
I.~Moerdijk and J.~van Oosten, \emph{Sets, Models and Proofs}.\hskip 1em plus
  0.5em minus 0.4em\relax Springer Cham, 2018.

\bibitem{Menhir}
\BIBentryALTinterwordspacing
F.~Pottier and Y.~R\'{e}gis-Gianas, ``What is {Menhir}?'' 2021. [Online].
  Available: \url{https://gallium.inria.fr/~fpottier/menhir/}
\BIBentrySTDinterwordspacing

\bibitem{jls3}
J.~Gosling, B.~Joy, G.~Steele, and G.~Bracha, \emph{The Java Language
  Specification, Third Edition}.\hskip 1em plus 0.5em minus 0.4em\relax
  Prentice Hall, June 2005.

\bibitem{Gusfield1997}
D.~Gusfield, \emph{Algorithms on Strings, Trees, and Sequences: Computer
  Science and Computational Biology}.\hskip 1em plus 0.5em minus 0.4em\relax
  Cambridge University Press, 1997.

\bibitem{Bille2005}
P.~Bille, ``A survey on tree edit distance and related problems,''
  \emph{Theoretical Computer Science}, vol. 337, no. 1-3, pp. 217--239, Jun.
  2005.

\bibitem{ZhangShasha1989}
K.~Zhang and D.~Shasha, ``Simple fast algorithms for the editing distance
  between trees and related problems,'' \emph{SIAM Journal on Computing},
  vol.~18, no.~6, pp. 1245--1262, 1989.

\bibitem{Imediff2003}
\BIBentryALTinterwordspacing
J.~Elonen, ``Imediff - an interactive fullscreen merge tool for diff2/3,''
  2003. [Online]. Available: \url{https://github.com/osamuaoki/imediff/}
\BIBentrySTDinterwordspacing

\bibitem{CavalcantiEtAl2017}
\BIBentryALTinterwordspacing
G.~Cavalcanti, P.~Borba, and P.~Accioly, ``Evaluating and improving
  semistructured merge,'' \emph{Proc. ACM Program. Lang.}, vol.~1, no. OOPSLA,
  pp. 59:1--59:27, Oct. 2017. [Online]. Available:
  \url{http://doi.acm.org/10.1145/3133883}
\BIBentrySTDinterwordspacing

\bibitem{ZhuEtAl2019}
F.~Zhu, F.~He, and Q.~Yu, ``Enhancing precision of structured merge by proper
  tree matching,'' in \emph{2019 IEEE/ACM 41st International Conference on
  Software Engineering: Companion Proceedings (ICSE-Companion)}, 2019, pp.
  286--287.

\bibitem{ApelEtAl2011}
\BIBentryALTinterwordspacing
S.~Apel, J.~Liebig, B.~Brandl, C.~Lengauer, and C.~K\"{a}stner,
  ``Semistructured merge: Rethinking merge in revision control systems,'' in
  \emph{Proceedings of the 19th ACM SIGSOFT Symposium and the 13th European
  Conference on Foundations of Software Engineering}, ser. ESEC/FSE '11.\hskip
  1em plus 0.5em minus 0.4em\relax New York, NY, USA: ACM, 2011, pp. 190--200.
  [Online]. Available: \url{http://doi.acm.org/10.1145/2025113.2025141}
\BIBentrySTDinterwordspacing

\bibitem{ZhuEtAl2018}
\BIBentryALTinterwordspacing
F.~Zhu and F.~He, ``Conflict resolution for structured merge via version space
  algebra,'' \emph{Proc. ACM Program. Lang.}, vol.~2, no. OOPSLA, pp.
  166:1--166:25, Oct. 2018. [Online]. Available:
  \url{http://doi.acm.org/10.1145/3276536}
\BIBentrySTDinterwordspacing

\bibitem{FalleriEtAl2014}
J.-R. Falleri, F.~Morandat, X.~Blanc, M.~Martinez, and M.~Monperrus,
  ``Fine-grained and accurate source code differencing,'' in \emph{Proceedings
  of the 29th ACM/IEEE International Conference on Automated Software
  Engineering}, ser. ASE '14.\hskip 1em plus 0.5em minus 0.4em\relax New York,
  NY, USA: ACM, 2014, pp. 313--324.

\bibitem{ChawatheEtAl97}
S.~S. Chawathe and H.~Garcia-Molina, ``Meaningful change detection in
  structured data,'' in \emph{Proceedings of the ACM SIGMOD International
  Conference on Management of Data}, Tuscon, Arizona, May 1997, pp. 26--37.

\bibitem{Spoon}
\BIBentryALTinterwordspacing
R.~Pawlak, M.~Monperrus, N.~Petitprez, C.~Noguera, and L.~Seinturier, ``{Spoon:
  A Library for Implementing Analyses and Transformations of Java Source
  Code},'' \emph{{Software: Practice and Experience}}, vol.~46, pp. 1155--1179,
  2015. [Online]. Available:
  \url{https://hal.archives-ouvertes.fr/hal-01078532/document}
\BIBentrySTDinterwordspacing

\bibitem{Miraldo2020}
V.~C. Miraldo, ``Type-safe generic differencing of mutually recursive
  families,'' Ph.D. dissertation, Utrecht University, October 2020.

\bibitem{GhiottoEtAl2018}
G.~Ghiotto, L.~Murta, M.~Barros, and A.~van~der Hoek, ``On the nature of merge
  conflicts: A study of 2,731 open source java projects hosted by {GitHub},''
  \emph{IEEE Transactions on Software Engineering}, vol.~46, no.~8, pp.
  892--915, 2020.

\bibitem{DeepMerge2023}
\BIBentryALTinterwordspacing
E.~Dinella, T.~Mytkowicz, A.~Svyatkovskiy, C.~Bird, M.~Naik, and S.~Lahiri, ``{
  DeepMerge: Learning to Merge Programs },'' \emph{IEEE Transactions on
  Software Engineering}, vol.~49, no.~04, pp. 1599--1614, Apr. 2023. [Online].
  Available: \url{https://doi.ieeecomputersociety.org/10.1109/TSE.2022.3183955}
\BIBentrySTDinterwordspacing

\bibitem{MahmoudiEtAl2019}
M.~Mahmoudi, S.~Nadi, and N.~Tsantalis, ``Are refactorings to blame? an
  empirical study of refactorings in merge conflicts,'' in \emph{Proceedings of
  the IEEE 26th International Conference on Software Analysis, Evolution and
  Reengineering}, ser. SANER '19, Feb 2019, pp. 151--162.

\bibitem{TsantalisEtAl2022}
N.~Tsantalis, A.~Ketkar, and D.~Dig, ``Refactoringminer 2.0,'' \emph{IEEE
  Transactions on Software Engineering}, vol.~48, no.~3, pp. 930--950, 2022.

\bibitem{EllisEtAl2023}
M.~Ellis, S.~Nadi, and D.~Dig, ``Operation-based refactoring-aware merging: An
  empirical evaluation,'' \emph{IEEE Transactions on Software Engineering},
  vol.~49, no.~4, pp. 2698--2721, 2023.

\bibitem{Mens2002Survey}
T.~{Mens}, ``A state-of-the-art survey on software merging,'' \emph{IEEE
  Transactions on Software Engineering}, vol.~28, no.~5, pp. 449--462, May
  2002.

\bibitem{Westfechtel1991}
\BIBentryALTinterwordspacing
B.~Westfechtel, ``Structure-oriented merging of revisions of software
  documents,'' in \emph{Proceedings of the 3rd International Workshop on
  Software Configuration Management}, ser. SCM '91.\hskip 1em plus 0.5em minus
  0.4em\relax New York, NY, USA: ACM, 1991, pp. 68--79. [Online]. Available:
  \url{http://doi.acm.org/10.1145/111062.111071}
\BIBentrySTDinterwordspacing

\bibitem{Buffenbarger1995}
J.~Buffenbarger, ``Syntactic software merging,'' in \emph{Software
  Configuration Management}, J.~Estublier, Ed.\hskip 1em plus 0.5em minus
  0.4em\relax Berlin, Heidelberg: Springer Berlin Heidelberg, 1995, pp.
  153--172.

\bibitem{MehdiEtAl2014}
\BIBentryALTinterwordspacing
A.-N. Mehdi, P.~Urso, and F.~Charoy, ``Evaluating software merge quality,'' in
  \emph{Proceedings of the 18th International Conference on Evaluation and
  Assessment in Software Engineering}, ser. EASE '14.\hskip 1em plus 0.5em
  minus 0.4em\relax New York, NY, USA: ACM, 2014, pp. 9:1--9:10. [Online].
  Available: \url{http://doi.acm.org/10.1145/2601248.2601275}
\BIBentrySTDinterwordspacing

\bibitem{AsenovEtAl2017}
D.~Asenov, B.~Guenat, P.~M{\"u}ller, and M.~Otth, ``Precise version control of
  trees with line-based version control systems,'' in \emph{Fundamental
  Approaches to Software Engineering}, M.~Huisman and J.~Rubin, Eds.\hskip 1em
  plus 0.5em minus 0.4em\relax Berlin, Heidelberg: Springer Berlin Heidelberg,
  2017, pp. 152--169.

\bibitem{SousaEtAl2018}
\BIBentryALTinterwordspacing
M.~Sousa, I.~Dillig, and S.~K. Lahiri, ``Verified three-way program merge,''
  \emph{Proc. ACM Program. Lang.}, vol.~2, no. OOPSLA, pp. 165:1--165:29, Oct.
  2018. [Online]. Available: \url{http://doi.acm.org/10.1145/3276535}
\BIBentrySTDinterwordspacing

\bibitem{DiasEtAl2015}
\BIBentryALTinterwordspacing
M.~Dias, G.~Polito, D.~Cassou, and S.~Ducasse, ``Deltaimpactfinder: Assessing
  semantic merge conflicts with dependency analysis,'' in \emph{Proceedings of
  the International Workshop on Smalltalk Technologies}, ser. IWST '15.\hskip
  1em plus 0.5em minus 0.4em\relax New York, NY, USA: ACM, 2015, pp. 8:1--8:6.
  [Online]. Available: \url{http://doi.acm.org/10.1145/2811237.2811299}
\BIBentrySTDinterwordspacing

\bibitem{NguyenEtAl2015}
\BIBentryALTinterwordspacing
H.~V. Nguyen, M.~H. Nguyen, S.~C. Dang, C.~K\"{a}stner, and T.~N. Nguyen,
  ``Detecting semantic merge conflicts with variability-aware execution,'' in
  \emph{Proceedings of the 2015 10th Joint Meeting on Foundations of Software
  Engineering}, ser. ESEC/FSE 2015.\hskip 1em plus 0.5em minus 0.4em\relax New
  York, NY, USA: ACM, 2015, pp. 926--929. [Online]. Available:
  \url{http://doi.acm.org/10.1145/2786805.2803208}
\BIBentrySTDinterwordspacing

\bibitem{HashimotoMori2008}
M.~Hashimoto and A.~Mori, ``{Diff/TS}: A tool for fine-grained structural
  change analysis,'' in \emph{Proceedings of the 15th Working Conference on
  Reverse Engineering}, ser. WCRE '08.\hskip 1em plus 0.5em minus 0.4em\relax
  Washington, DC, USA: IEEE Computer Society, 2008, pp. 279--288.

\bibitem{FluriEtAl2007}
B.~Fluri, M.~W{\"{u}}rsch, M.~Pinzger, and H.~Gall, ``Change distilling: Tree
  differencing for fine-grained source code change extraction,'' \emph{IEEE
  Transactions on Software Engineering}, vol.~33, no.~11, pp. 725--743, 2007.

\end{thebibliography}

\end{document}